\documentclass[preprint2]{aastex}
\usepackage{epsf,graphicx}

\begin{document}

\def\la{\mathrel{\hbox{\rlap{\hbox{\lower4pt\hbox{$\sim$}}}\hbox{$<$}}}}
\def\ga{\mathrel{\hbox{\rlap{\hbox{\lower4pt\hbox{$\sim$}}}\hbox{$>$}}}}

\font\sevenrm=cmr7
\def\HeI{He~{\sevenrm I}}
\def\HeII{He~{\sevenrm II}}
\def\OI{O~{\sevenrm I}}
\def\OIII{[O~{\sevenrm III}]}
\def\CaII{Ca~{\sevenrm II}}
\def\FeI{Fe~{\sevenrm I}}
\def\FeII{Fe~{\sevenrm II}}
\def\NaI{Na~{\sevenrm I}}
\def\FeIIf{[Fe~{\sevenrm II}]}
\def\OII{[O~{\sevenrm II}]}
\def\SII{[S~{\sevenrm II}]}
\def\SIII{[S~{\sevenrm III}]}
\def\SVIII{[S~{\sevenrm VIII}]}
\def\SIX{[S~{\sevenrm IX}]}
\def\CI{[C~{\sevenrm I}]}
\def\NI{[N~{\sevenrm I}]}
\def\SiVI{[Si~{\sevenrm VI}]}
\def\SiX{[Si~{\sevenrm X}]}
\def\PII{[P~{\sevenrm II}]}
\def\CaVIII{[Ca~{\sevenrm VIII}]}
\def\FeXI{[Fe~{\sevenrm XI}]}
\def\NII{[N~{\sevenrm II}]}


\newdimen\digitwidth
\setbox0=\hbox{-.}
\digitwidth=\wd0
\catcode `@=\active
\def@{\kern\digitwidth}

\title{The Near-Infrared Broad Emission Line Region of Active Galactic Nuclei -- I. The Observations}

\author{Hermine Landt\altaffilmark{1}}
\affil{Harvard-Smithsonian Center for Astrophysics, 60 Garden Street, 
Cambridge, MA 02138.}

\author{Misty C. Bentz\altaffilmark{1}}
\affil{Department of Astronomy, The Ohio State University, 
140 West 18th Avenue, Columbus, OH 43210.}

\author{Martin J. Ward\altaffilmark{1}}
\affil{Department of Physics, University of Durham, South Road, 
Durham, DH1 3LE, UK.}

\author{Martin Elvis\altaffilmark{1}}
\affil{Harvard-Smithsonian Center for Astrophysics, 60 Garden Street, 
Cambridge, MA 02138.}

\author{Bradley M. Peterson}
\affil{Department of Astronomy, The Ohio State University, 
140 West 18th Avenue, Columbus, OH 43210.}

\author{Kirk T. Korista}
\affil{Department of Physics, Western Michigan University, 
1903 W. Michigan Avenue, Kalamazoo, MI 49008.}

\author{Margarita Karovska}
\affil{Harvard-Smithsonian Center for Astrophysics, 60 Garden Street, 
Cambridge, MA 02138.}

\altaffiltext{1}{Visiting Astronomer at the Infrared Telescope
Facility, which is operated by the University of Hawaii under
Cooperative Agreement no. NCC 5-538 with the National Aeronautics and
Space Administration, Science Mission Directorate, Planetary Astronomy
Program.}

\begin{abstract}

We present high quality (high signal-to-noise ratio and moderate
spectral resolution) near-infrared (near-IR) spectroscopic
observations of 23 well-known broad-emission line active galactic
nuclei (AGN). Additionally, we obtained simultaneous (within two
months) optical spectroscopy of similar quality. The near-IR broad
emission line spectrum of AGN is dominated by permitted transitions of
hydrogen, helium, oxygen, and calcium, and by the rich spectrum of
singly-ionized iron. In this paper we present the spectra, line
identifications and measurements, and address briefly some of the
important issues regarding the physics of AGN broad emission line
regions. In particular, we investigate the excitation mechanism of
neutral oxygen and confront for the first time theoretical predictions
of the near-IR iron emission spectrum with observations.

\end{abstract}


\section{Introduction}

Strangely, the near-infrared (near-IR) spectra of broad-emission line
active galactic nuclei (type 1 AGN) have been mostly ignored in this
new era of infrared spectroscopy. The broad ($\sim 1\%-5\%$ the speed
of light) emission lines observed in the optical and ultraviolet (UV)
spectra of these type 1 AGN are the most direct tracers of the
activity and immediate environment of supermassive black
holes. However, despite 30 years of intensive spectrophotometric
studies the geometry and kinematics of the broad emission line region
(BELR) remains ill-defined. It is not clear whether the BELR gas has a
spherical or disk-like distribution, and whether it consists of a
large number of discrete clouds or is part of an outflowing,
continuous gas distribution such as an accretion disk wind \citep[see,
e.g., review by][]{Sul00}. The knowledge of the BELR geometry and
kinematics is essential not only to our understanding of the relation
between different types of AGN but also to studies that use the broad
emission line widths to estimate black holes masses. These estimates
would not be meaningful if the BELR was in outflow, and, if the gas
was gravitationally bound but distributed in a disk, the resulting
values would be underestimated for sources viewed face-on.

Our current, limited knowledge of the BELR physical conditions and
scales was gained primarily through the application of sophisticated
photoionization models to the observed emission line intensities and
ratios \citep[see, e.g., review by][]{Ferl03} and through
reverberation mapping studies of the (correlated) continuum and line
variability \citep[see, e.g., review by][]{Pet93}. Both of these
methods have relied so far on spectroscopic observations at optical
and UV frequencies. Our goal is to extend these studies to the near-IR
wavelengths.

Infrared broad emission lines offer several advantages: (i) since each
emission line is formed most efficiently at a particular density and
distance from the ionizing source \citep{Bal95}, it is important to
incorporate in photoionization and reverberation mapping studies
multiple emission lines to map the entire BELR. In this respect, the
near-IR broad emission lines are a vital addition to the lines
observed at optical and UV frequencies; (ii) the near-IR broad
emission lines trace mainly the low-ionization line (LIL) region, a
region believed to have extreme properties such as very high densities
($n>10^{11}$ cm$^{-3}$) and a disk-like structure \citep{Coll88,
Marz96}. However, these properties were derived mainly based on
observations of the two strongest Balmer lines, H$\alpha$ and
H$\beta$, and need additional verification; (iii) the near-IR broad
emission lines are little affected by reddening and, therefore, can
yield information on the amount of dust present in the BELR when
compared to the optical Balmer lines; and (iv) coordinated
spectroscopic observations at different wavelengths, which are
necessary to cover simultaneously a large number of emission lines in
the variable AGN, are easiest to realize at optical and near-IR
frequencies.

An additional strong motivation for our observing program was to
identify nearby AGN with near-IR broad emission lines sufficiently
luminous to allow imaging with the newly commissioned and future
near-IR interferometers, such as, e.g., {\sl VLTI} and {\sl
Ohana}. The combination of an interferometric map and a
reverberation-based velocity-delay map can be expected to yield the
BELR structure and kinematics with almost no ambiguity
\citep[e.g.,][]{Gal01}. In the longer term we will also use these
studies to measure the metric of the Universe by purely geometrical
means, following \citet{Elvis02}. As these authors pointed out, by
using reverberation mapping to measure the linear size and the
three-dimensional structure of the BELR and an interferometer to
measure its angular size it is possible to solve the triangle (as in
stellar parallax), giving the angular diameter distance to the
quasar. Mapping out angular diameter distance versus redshift gives
the metric, and quasars are common and bright up to high redshifts
($z\sim 3$), where the difference between $\Lambda=0.7$ and
$\Lambda=0.0$ reaches 40\%.

This paper presents near-IR spectra of a relatively large sample (23
sources) of type 1 AGN and is the first to include complementary
contemporaneous optical spectra to allow the evaluation of commonly
measured broad emission line ratios. We concentrate here on the
detailed description of our near-IR and optical spectroscopic
observations (Section 2), line identifications (Section 3), and flux
and velocity width measurements of the most important near-IR and
optical broad emission lines (Section 4). In Section 5 we briefly
discuss our results in the context of the general BELR physics. The
detailed theoretical analysis of these data will be presented in a
series of forthcoming papers.

\section {The Observations}

We selected for observations well-known AGN with broad emission lines
at redshifts such that most of the strong Paschen hydrogen lines fall
in the near-IR atmospheric windows. In addition we required that our
sources are bright enough in the near-IR (magnitudes $J\la14$ mag) to
allow us to obtain high signal-to-noise ratio (S/N $\sim 100$) spectra
in a reasonable amount of exposure time. As of June 2006 our observed
sample comprises 23 sources. In Table \ref{general} we summarize their
general properties.

We observed with the NASA Infrared Telescope Facility (IRTF), a 3 m
telescope on Mauna Kea, Hawai'i. We used the SpeX spectrograph
\citep{Ray03} set in the short cross-dispersed mode (SXD, $0.8-2.4$
$\mu$m) and equipped with the $0.8\times15''$ slit. This set-up gives
an average spectral resolution of full width at half maximum (FWHM)
$\sim 400$ km s$^{-1}$. Before and/or after each source, we observed a
nearby (in position and airmass) A0V star with well-known $B$ and $V$
magnitudes.  These stars were used to correct the source spectra for
telluric absorption \citep{Vacca03} and for flux calibration. Flats
and arcs were taken with each source/telluric standard star pair. The
data were collected during three observing runs in the period of 2004
May - 2006 June. We had clear skies and seeing in the range of $\sim
0.7 - 1''$ for all nights with the exception of 2006 January 9 and 10
which were affected by clouds and had poor seeing of $\sim 1.5''$. In
Table \ref{irtfobslog} we give the journal of observations.

The SpeX data were reduced using Spextool, an Interactive Data
Language (IDL)-based software package developed for SpeX users
\citep{Cush04}. Spextool carries out all the procedures necessary to
produce fully reduced spectra. This includes preparation of
calibration frames, processing and extraction of spectra from science
frames, wavelength calibration of spectra, telluric correction and
flux-calibration of spectra, and merging of the different orders into
a single, continuous spectrum. The extraction width was adjusted
interactively for each source to include all the flux in the spectral
trace. The final spectra were corrected for Galactic extinction using
the IRAF task \mbox{\sl onedspec.deredden} with input $A_{\rm V}$
values derived from Galactic hydrogen column densities published by
\citet{DL90}. The results are shown in Fig. \ref{IRTFspectra}, where
we plot for each source the spectrum with the highest S/N. The
spectral fits-files are available for download from the journal as a
tar-file.

For all our sources we also obtained contemporaneous (within two
months) optical spectroscopy with the FAST spectrograph \citep{Fast98}
at the Tillinghast 1.5 m telescope on Mt. Hopkins, Arizona, in
queue-observing mode. We used the 300 l/mm grating and a $3''$
slit. This set-up gives a spectral coverage of $\sim 3720 - 7515$
\AA~and an average spectral resolution of FWHM $\sim 330$ km
s$^{-1}$. The slit was rotated to the parallactic angle for all
observations. In Table \ref{flwoobslog} we give the journal of
observations.

The FAST data were reduced using standard routines from the IRAF
software package. In particular the 2-dimensional spectral files were
trimmed, overscan- and bias-subtracted, normalized, rectified and
wavelength-calibrated. Subsequently the spectrum was extracted using a
width that included all the flux in the trace and the 1-dimensional
spectral file flux-calibrated using photometric standard stars
observed the same night. The final spectra were corrected for Galactic
extinction (see above). The spectral fits-files are available for
download from the journal as a tar-file.

\section{The Emission Line Identifications}

\begin{figure*}
\centerline{
\includegraphics[clip=true,bb=0 450 590 715,scale=1.0]{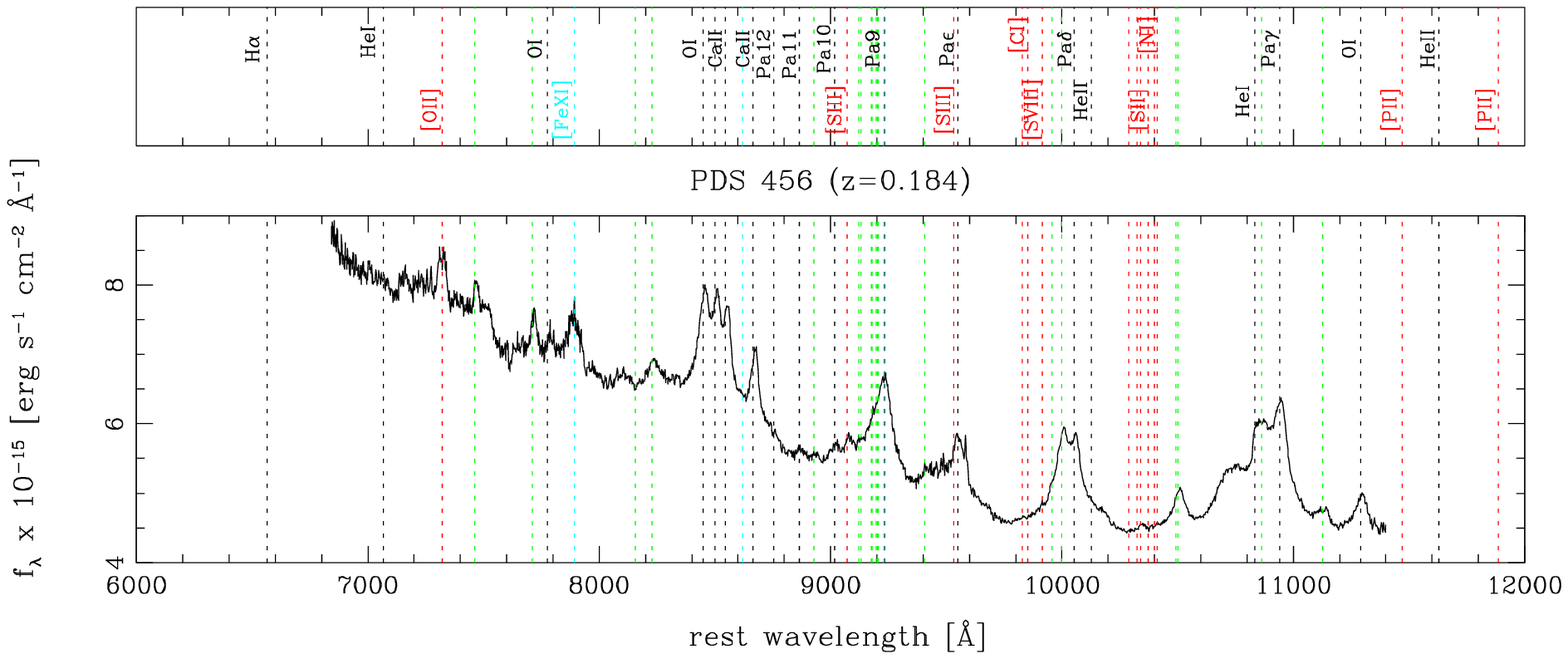}
}
\vspace*{-1cm}
\centerline{
\includegraphics[clip=true,bb=0 450 590 715,scale=1.0]{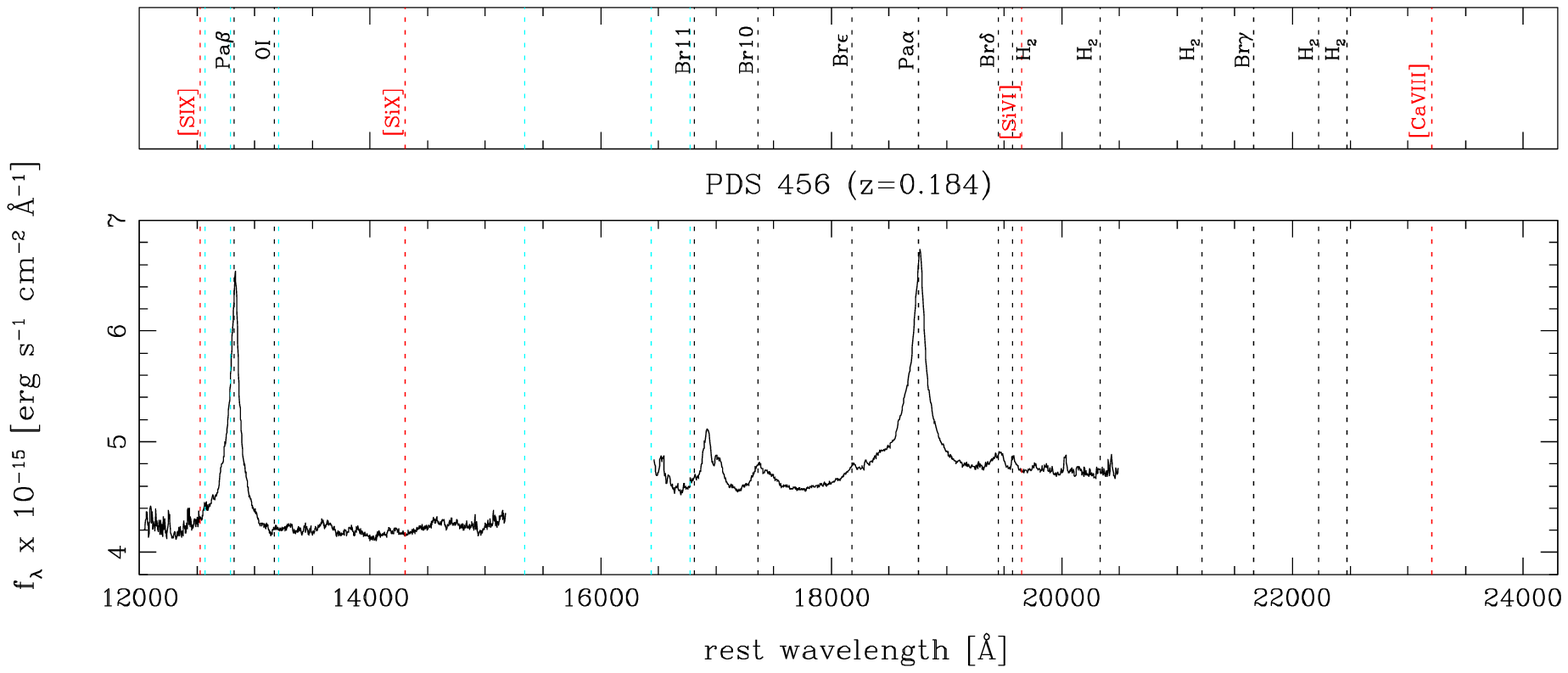}
}
\caption{\label{IRTFspectra} IRTF SpeX near-infrared spectrum of PDS
456 shown as observed flux versus rest frame wavelength. Identified
emission lines are marked by dotted lines and labeled: black,
permitted transitions; green, permitted \FeII~multiplets (not
labeled); red, forbidden transitions; cyan, forbidden Fe transitions
with only those higher than [\FeII] labeled. [{\it The spectral plots
for all sources are available in colour in the electronic edition of
the journal. The spectra used to create these plots are available for
download as a tar-file.}]}
\end{figure*}

We have identified in our near-IR spectra all observed emission lines,
both permitted and forbidden transitions (black and red dotted lines,
respectively, in Fig. \ref{IRTFspectra}), as well as the rich
permitted and forbidden transitions of iron, in particular of \FeII~
(green and cyan dotted lines, respectively, in
Fig. \ref{IRTFspectra}). The line identifications were gathered from
the literature \citep[e.g.,][]{Mor88, Ost90, Oliva94, Thom95, Ost96,
Rudy00, Van01, Rod02a, Sig03, Veron04, Rif06, Maz07} and the vacuum
rest-frame wavelengths were taken from the Atomic Line List
v2.04\footnote{The Atomic Line List is a compilation by Peter van Hoof
and is hosted by the Department of Physics and Astronomy at the
University of Kentucky (see
http://www.pa.uky.edu/$\sim$peter/atomic/)}, with the exception of
those of the permitted \FeII~multiplets, which were taken from
\citet{Sig03}, and those of molecular hydrogen, which were taken from
\citet{Rif06}.

In Table \ref{ident} we summarize the identified emission lines in the
near-IR frequency range. For six prominent features we could not find
a suitable identification using the \FeII~multiplet list of
\citet{Sig03}, but profile comparisons suggest that they are iron
emission lines. We discuss these further in Section \ref{noID}. We
note that an extensive list of optical emission lines can be found in
\citet{Van01}.

\section{The Measurements} \label{measurements}

We have measured the properties of the most important broad emission
lines in the spectra of our sources. In order to obtain accurate
measures, we have extracted ``pure'' broad line profiles which
involved the following steps: (i) estimation and subtraction of the
continuum; (ii) subtraction of blended emission lines of other
elements; (iii) removal of the superposed narrow emission line
component; and (iv) modeling and subtraction of the
\FeII~emission. These steps are discussed in detail in the following
subsections.

On the extracted ``pure'' broad emission line profile we measured the
flux, FWHM, and shift. The flux was measured with the IRAF task {\sl
onedspec.splot} by integrating the emission in the line, i.e., we did
not assume a specific profile. The shift was measured as the velocity
difference between the theoretical (vacuum) and observed wavelengths
of the emission line center, where the emission line center was
measured as the center of the width at half maximum. The instrumental
width has not been removed from the FWHM values. All measurements were
performed on the redshift-corrected spectrum (i.e., in the object's
rest-frame). We note that the redshifts were taken from the NASA/IPAC
Extragalactic Database (NED) and, therefore, are likely to be
inhomogeneous in their definition of the object's rest-frame.

\subsection{The Continuum}

We have estimated for each broad emission line a local continuum. This
task involved, firstly, defining the limits of the wings of the broad
emission line, and, secondly, finding line-free continuum regions to
each side of the broad emission line large enough to allow for a
reliable fit. Since the emission line widths differ widely in our
sources we did not attempt to define global line-free continuum
regions but rather adjusted the fitted continuum regions individually
for each object (and line) and chose them to be as close as possible
to the wings and as large as possible. The local continuum was then
fitted with a straight line in $f_\lambda$ vs. $\lambda$ space and
subsequently subtracted.

\subsection{The Blends} \label{blends}

We have cleaned the spectrum of narrow emission lines of other
elements within $\sim 12000$ km s$^{-1}$ (approximately the largest
half-width at zero intensity observed) of the broad emission line's
expected rest-frame wavelength by modeling the features with Gaussians
and subtracting them from the spectrum using the IRAF task {\sl
onedspec.splot}. When required, narrow emission lines were first
deblended using the option 'd' of this task before removal. In the
case of the oxygen doublet \OIII~$\lambda \lambda 4959, 5007$ we did
not assume a Gaussian profile but isolated the emission lines from the
spectrum by interpolating the H$\beta$ profile under them.

In the case of the near-IR broad emission line Pa$\epsilon$, which is
heavily blended with the \SIII~$\lambda 9531$ narrow emission line, we
used either the Pa$\alpha$ or Pa$\beta$ profile (whichever had the
higher S/N ratio; mostly this was Pa$\beta$) as a template. Both these
emission lines are excellent templates since they are observed
relatively free of strong blends. We scaled the Paschen template
profile to the peak intensity of Pa$\epsilon$ and subtracted it from
the blend. From the result we isolated the \SIII~profile and in turn
subtracted this from the total blend to recover, at least in part, the
original broad emission line.

We have also used the Paschen template technique described above for
the broad emission line blends Pa$\gamma$ and \HeI~1.0830 $\mu$m, and
Pa$\delta$ and \HeII~1.0124 $\mu$m. In the case of Pa$\gamma$ and
\HeI, we first subtracted the scaled template profile corresponding to
the broad emission line with the highest peak, since the peak of the
weaker broad emission line is severely contaminated by flux from the
wings of the stronger component.

Similarly, we have used the profile of the (unblended) \OI~1.1287
$\mu$m broad emission line as a template in order to deblend the
\OI~$\lambda 8446$ broad emission line from the \CaII~triplet. The
only exceptions were PDS 456 and H 1821$+$643, where we used the
Pa$\beta$ profile instead. In PDS 456, the \OI~1.1287 $\mu$m profile
is much narrower than that of \OI~$\lambda 8446$, and in H 1821$+$643,
\OI~1.1287 $\mu$m appears to be blue-shifted relative to \OI~$\lambda
8446$ by $\sim 1400$ km s$^{-1}$.

\subsection{The Narrow Component} \label{narrow}

The task {\sl onedspec.splot} was also used to model the narrow
component with a Gaussian and remove it from the broad emission line
profile. The narrow line profile appears inflected (i.e., it is
clearly distinguished from the broad line profile; see, e.g., Mrk 290
in Fig. \ref{PaAPaB}) in 11/23 sources. In these cases the modeling
and removal process is straightforward. In another two sources,
namely, PG 0844$+$349 and Ark 120, the broad line profiles (with the
exception of those of \HeI~1.0830 $\mu$m) had a broad top, either flat
or double-peaked. We interpreted this as indication that the
contribution from the narrow line component is negligible and left the
observed profiles unchanged.

The broad emission lines of the remaining 10/23 sources were peaked,
but the transition between broad and narrow component was not
perceptible. In these cases we estimated the contribution of the
narrow component to the total profile by fitting to its top part a
Gaussian with FWHM equal to that of the narrow emission line
\OIII~$\lambda 5007$. This method assumes that the FWHM of
\OIII~$\lambda 5007$ is representative of the narrow emission line
region (NELR) and subtracts the largest possible flux contribution
from this region. 

No attempt was made to correct the broad line profiles of IRAS
1750$+$508 and PDS 456 for the presence of a narrow line
component. The \OIII~emission lines of IRAS 1750$+$508 show a very
strong blue asymmetry, and those of PDS 456 are very weak and strongly
blended with (optical) \FeII.

\subsection{The \FeII~Subtraction} \label{ironsub}

Most near-IR broad emission lines suffer from moderate to strong
blending with \FeII~emission features. The template subtraction
process described in Section \ref{blends} isolates these
automatically. However, it is necessary to judge if the isolated
profile indeed corresponds to \FeII~or rather belongs, in part or
entirely, to a different, neighboring emission feature (e.g., to
\HeII~1.0124 $\mu$m in the case of \FeII~1.0132 $\mu$m) or the wing of
the broad emission line in question. To determine this, we compared
the isolated profile to the profiles of two near-IR \FeII~emission
features which are relatively strong and observed unblended, namely,
\FeII~1.0502 $\mu$m and \FeII~1.1126 $\mu$m. The \FeII~1.0502 $\mu$m
emission feature is in fact a blend of \FeII~1.0491 $\mu$m and
\FeII~1.0502 $\mu$m. However, the former is expected to be weaker than
the latter by a factor of $\sim 7$ \citep{Sig03}.

In the optical frequency range the continuum around H$\beta$,
\HeII~$\lambda 4686$, and H$\gamma$ is contaminated by numerous
\FeII~multiplets which blend together to form a pseudo-continuum. We
first subtracted this optical \FeII~emission before fitting a local
continuum. For this purpose we used the template based on the optical
spectrum of I Zw 1 published by \citet{Veron04} and available in
electronic format. The method generally used to subtract the
\FeII~emission from optical spectra was first introduced by
\citet{Bor92}. It consists of creating a spectral sequence by
broadening (by convolution with Gaussians) and scaling of the
\FeII~template, which is subsequently packed together into a
three-dimensional cube. This cube is then subtracted from a cube
consisting in all three dimensions of the object's spectrum.

In contrast to \citet{Bor92}, in our cases we found it difficult to
decide by eye unambiguously which pair of width and strength of the
\FeII~template gave the cleanest subtraction. Whereas it was rather
clear which was the most appropriate strength, changing the width of
the \FeII~template (at this strength) by as much as several 1000 km
s$^{-1}$ gave only slightly different results. This is due to the fact
that the \FeII~template ``smears out'' at larger widths, which leads
to essentially subtracting a straight line over large continuum
regions. (However, we note that the resulting differences were more
pronounced at smaller widths, and in principle a minimum width could
be determined.) This degeneracy was also noted by \citet{Ves05}.
Therefore, it seems necessary to constrain a priori the width of the
\FeII~template. We did so by using the widths measured for our
unblended near-IR iron emission lines \FeII~1.0502 $\mu$m or
\FeII~1.1126 $\mu$m, whichever had the higher S/N ratio (usually
\FeII~1.0502 $\mu$m). Three sources have near-IR \FeII~emission lines
narrower than the width of the \FeII~template of 1100 km s$^{-1}$. In
these cases we did not broaden the template. For two sources, namely,
H 1821$+$643 and NGC 5548, we do not observe near-IR \FeII~emission
(except a weak near-IR \FeII~emission 'hump'), but we do observe
(weak) optical \FeII~emission. In these cases we have not subtracted
an \FeII~template, but simply fit a local continuum using the spectral
ranges $\lambda 4435 - 4445$~\AA~and $\lambda 5075 - 5085$~\AA, where
\FeII~is expected to be very weak.

In many cases, and especially for strong iron emitters such as, e.g.,
PDS 456, no satisfactory optical \FeII~subtraction using the template
of \citet{Veron04} was possible. Strong emission residuals remained in
particular in the range $4440-4700$~\AA~and also around \FeII~$\lambda
4233$. This is most likely due to the fact that the flux ratios
between the individual \FeII~emission lines differ from object to
object, and thus between our sources and I Zw 1 \citep[see also,
e.g.,][]{Bor92}. Therefore, in order to not leave large amounts of
flux unaccounted for, we have re-added the corresponding, subtracted
\FeII~template to the continuum-subtracted spectrum, and, if present,
subtracted the isolated \HeII~$\lambda 4686$ profile, before measuring
the \FeII~emission in the range $\lambda 4440-4700$. The exception was
the source NGC 4151 for which we measured the optical \FeII~emission
directly in the subtracted template, since its H$\beta$ broad emission
line has very extended wings stretching out to relatively short
wavelengths. We note that measuring instead the flux of \FeII~$\lambda
4440-4700$ directly in the subtracted template can lead to
underestimating it by as much as $\sim 40\%$ and typically by $\sim
15\%$.

\subsection{Host Galaxy Absorption} \label{host}

Since our optical spectra were obtained through a rather large
aperture in objects with strong host galaxy contribution, the
H$\gamma$ and \HeI~$\lambda 5876$ emission lines are affected by
stellar absorption \citep[see also][]{Cren85}. In these cases we used
either the Pa$\alpha$ or Pa$\beta$ profile as a template to
interpolate the H$\gamma$ profile over the strongest absorption lines
(most notably G band at 4300~\AA), which, when they add together, can
appear rather wide.

The absorption lines contaminating \HeI~$\lambda 5876$ (e.g.,
\NaI~$\lambda 5896$) were instead fitted with Gaussians and
subtracted. We did not attempt to measure the \HeI~$\lambda 5876$
emission line in PDS 456 and Mrk 590. In both sources this line was
strongly absorbed, by telluric bands and by the host galaxy,
respectively.

\subsection{The Tables}

In Table \ref{lines} we list our results individually for each
measured broad emission line. Sources are omitted from the list if the
spectrum did not cover the line. If not the entire broad emission line
but more than half of it is covered by the spectrum, we list lower
limits on the flux. The fluxes of selected \FeII~emission lines are
listed in Table \ref{iriron} and are discussed further in Section
\ref{iron}. The fluxes of isolated narrow emission lines of other
elements are listed in \mbox{Table \ref{irnarrow}}.

The main sources of measurement errors on the listed broad emission
line parameters are the placement of the continuum and the setting of
the extension of an emission line. Both of these depend to a large
extent on the S/N of the data but also significantly on the subjective
decision of the person conducting the measurements. For these reasons
we have not attempted to derive individual errors for our
measurements. However, the S/N of our data is relatively high and we
estimate the error introduced by this quantity alone to be, e.g., a
few per cent for the broad emission line fluxes. Measurement errors
for the narrow emission line fluxes are due mainly to the
approximation with a Gaussian and the placement of the 'narrow
emission line continuum'. The Gaussian fit was adequate in the
majority of cases, but in some objects this approximation might not
account for all of the flux. In general we estimate that the listed
values are accurate to within $\sim 10\%$.

\section{The Near-IR Broad Emission Lines}

The electron densities in the BELR are sufficiently high ($n \ga 10^8$
cm$^{-3}$) that virtually all forbidden lines are collisionally
suppressed, therefore, almost only permitted transitions contribute to
this emission line region. We find that the near-IR spectra of our
sources are dominated by permitted emission lines from few elements,
namely, hydrogen, helium, oxygen, calcium, and iron. This is similar
to what is observed in the optical frequency range.

\subsection{The Hydrogen Emission Lines}

\begin{figure*}
\centerline{
\includegraphics[clip=true,bb=1 519 590 710,scale=1.0]{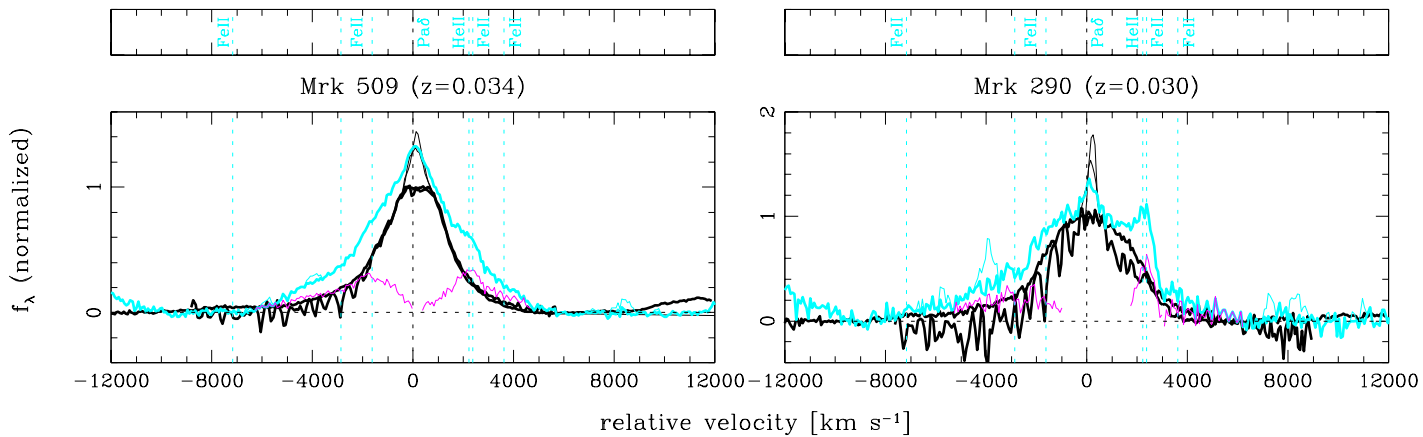}
}
\caption{\label{PaAPaBPaD}  IRTF SpeX profiles of the
emission line blend Pa$\delta$, \HeII~1.0124 $\mu$m, \FeII~$\lambda
9812$, \FeII~$\lambda 9956+9998$, \FeII~1.0132 $\mu$m, and
\FeII~1.0174 $\mu$m (cyan, thick lines) compared to the profiles of
Pa$\alpha$ and Pa$\beta$ (black). Profiles are shown in velocity space
relative to the expected rest frame wavelength indicated by the
vertical dotted lines (black for the Paschen lines, cyan for \FeII~and
\HeII). The profiles have been continuum-subtracted and normalized to
the same peak intensity (of the broad component). The subtracted
narrow emission line blends (cyan, thin lines) and the isolated
\FeII~and \HeII~profiles (magenta) are also shown. [{\it The spectral
plots for all sources are available in colour in the electronic
edition of the journal.}]}
\end{figure*}

\begin{figure*}
\centerline{
\includegraphics[clip=true,bb=1 519 590 710,scale=1.0]{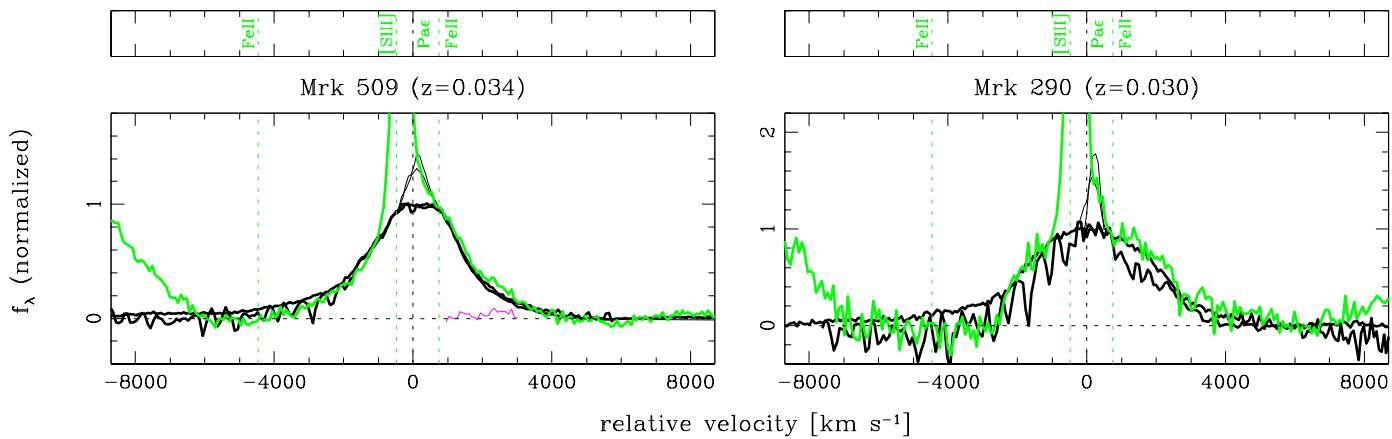}
}
\caption{\label{PaAPaBPaE}  IRTF SpeX profiles of the
emission line blend Pa$\epsilon$, \SIII~$\lambda 9531$, \FeII~$\lambda
9407$, and \FeII~$\lambda 9573$ (green, thick lines) compared to the
profiles of Pa$\alpha$ and Pa$\beta$ (black). Profiles are shown in
velocity space relative to the expected rest frame wavelength
indicated by the vertical dotted lines (black for the Paschen lines,
green for \FeII~and \SIII). The profiles have been
continuum-subtracted and normalized to the same peak intensity (of the
broad component). The isolated \FeII~profiles (magenta) are also
shown. [{\it The spectral plots for all sources are available in
colour in the electronic edition of the journal.}]}
\end{figure*}

\begin{figure*}
\centerline{
\includegraphics[clip=true,bb=1 545 590 710,scale=1.0]{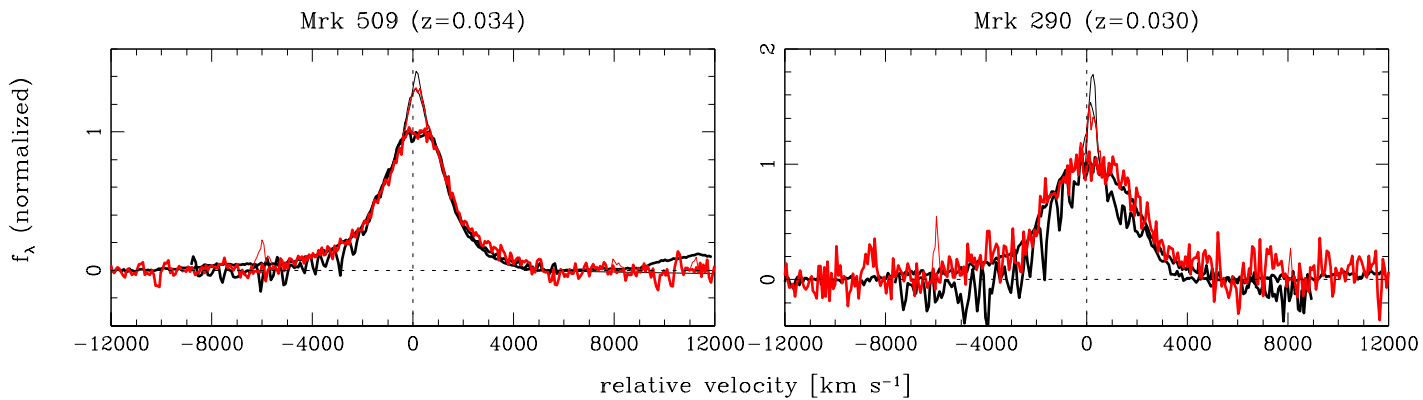}
}
\caption{\label{PaAPaBBrC}  IRTF SpeX profiles of the
broad emission line Br$\gamma$ (red) compared to the profiles of
Pa$\alpha$ and Pa$\beta$ (black) in velocity space relative to the
expected rest frame wavelength (thick lines). The profiles have been
continuum-subtracted and normalized to the same peak intensity (of the
broad component). The subtracted blends and narrow components are also
shown (thin lines). [{\it The spectral plots for all sources are
available in colour in the electronic edition of the journal.}]}
\end{figure*}

The observed near-IR hydrogen emission lines contain transitions from
the Paschen and Brackett series. We have measured 5/11 observed
Paschen emission lines, namely, Pa$\alpha$ to Pa$\epsilon$ (shown in
Figs. \ref{PaAPaB}-\ref{PaAPaBPaE}), and the strongest observed
Brackett emission line, namely, Br$\gamma$ (shown in
Fig. \ref{PaAPaBBrC}). Although seen in the majority of our sources
(21/23 objects), we have not measured the second strongest observed
Brackett emission line, namely, Br$\delta$, since it is strongly
affected by telluric absorption. In the optical spectra of our sources
we have measured the three strongest Balmer emission lines, namely,
H$\alpha$, H$\beta$, and H$\gamma$.

\subsubsection{The Paschen and Balmer Series}

\begin{figure*}
\centerline{
\includegraphics[clip=true,bb=1 545 590 710,scale=1.0]{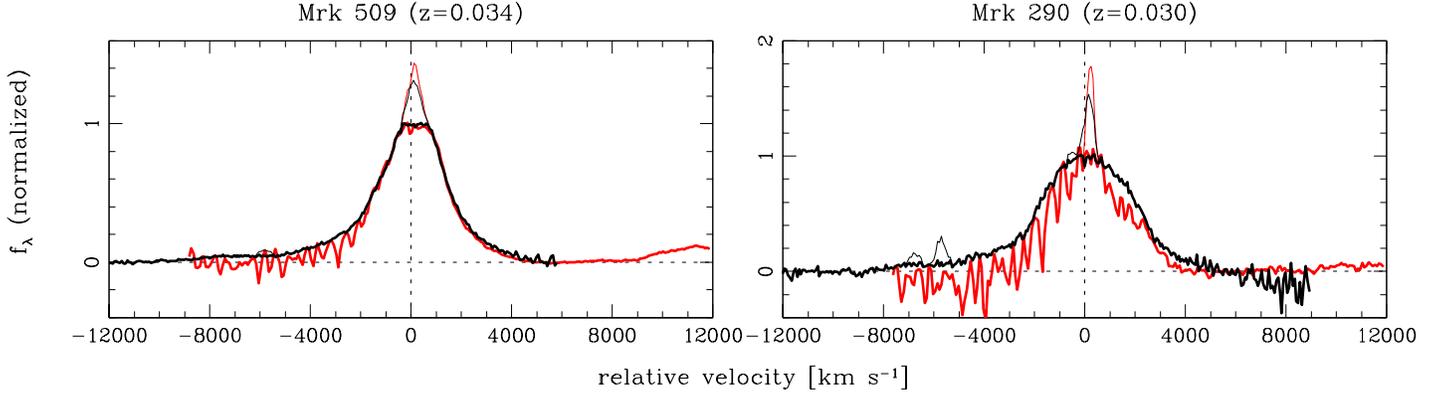}
}
\caption{\label{PaAPaB}  IRTF SpeX profiles of the broad
emission lines Pa$\alpha$ (red) and Pa$\beta$ (black) in velocity
space relative to the expected rest frame wavelength (thick
lines). The profiles have been continuum-subtracted and normalized to
the same peak intensity (of the broad component). The subtracted
blends and narrow components are also shown (thin lines). [{\it The
spectral plots for all sources are available in colour in the
electronic edition of the journal.}]}
\end{figure*}

The Pa$\alpha$ and Pa$\beta$ emission lines are not only two of the
strongest emission lines observed in the near-IR, but they are also
relatively free of blending features (only two, mostly weak narrow
emission lines contaminate the blue wing of Pa$\beta$, see
Fig. \ref{PaAPaB}). This makes them valuable tools to probe the
physical conditions of the BELR. The optical Balmer emission lines,
although stronger, suffer from blending with other features and, at
least to some degree, are expected to be affected by dust absorption.

\begin{figure*}
\centerline{ \includegraphics[scale=0.42]{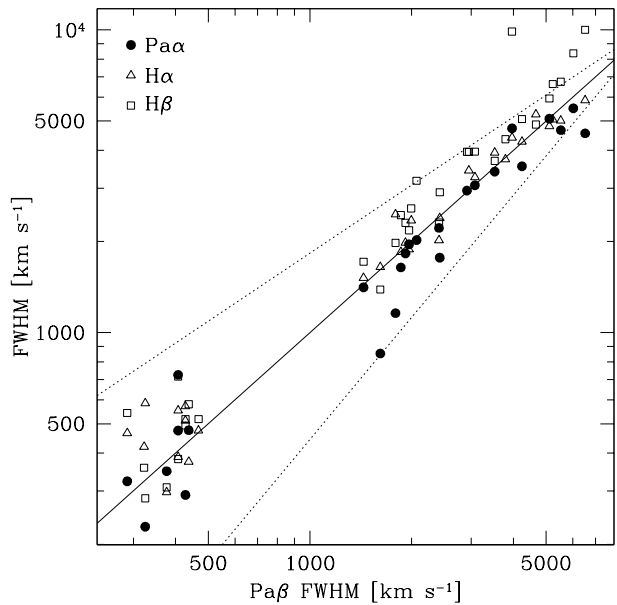} }
\caption{\label{PaBalFWHM}  The full width at half
maximum (FWHM) of the broad and narrow {\sl inflected} components
(upper right and lower left symbols, respectively) of Pa$\beta$
vs. Pa$\alpha$ (filled circles), H$\alpha$ (open triangles), and
H$\beta$ (open squares). The compared emission lines have similar
widths (solid line) within the error range given by the spectral
resolution (dotted lines). However, in several sources the broad
Pa$\alpha$ component suffers slightly from atmospheric absorption, and
there is a trend for the H$\beta$ broad component to be wider than
that of Pa$\beta$. The latter effect is most likely due to the
presence of the H$\beta$ ``red shelf''.}
\end{figure*}

Given this importance of Pa$\alpha$ and Pa$\beta$, it is worth
comparing their widths. We do this in Fig. \ref{PaBalFWHM} where we
plot separately their broad and narrow components (upper right and
lower left filled circles, respectively), the latter plotted only for
the cases where it is inflected. Taking the spectral resolution (FWHM
$\sim 350$ km s$^{-1}$) as an error range (dotted lines), as expected,
the widths of these two emission lines are similar. However, in
several sources we observe the Pa$\alpha$ emission line close to the
atmospheric absorption window and this is reflected in
Fig. \ref{PaBalFWHM}. Whereas the scatter for the narrow component is
evenly distributed, the scatter in the broad component measurements is
mainly towards smaller values for Pa$\alpha$ relative to Pa$\beta$. As
an example, the two most deviant points at the lower and upper end of
the broad component distribution correspond to Ark 564 and NGC 5548,
respectively. And, as can be seen from Fig. \ref{PaAPaB}, the
Pa$\alpha$ broad emission lines in these two sources suffers from
extreme atmospheric absorption.

All Paschen emission lines of higher order are strongly blended with
emission lines from other elements. In particular, Pa$\gamma$ is
strongly blended with \HeI~1.0830 $\mu$m and \FeII~1.0863 $\mu$m, and
its red wing also with \FeII~1.1126 $\mu$m (see
Fig. \ref{PaAPaBPaC}). The Pa$\delta$ emission line is heavily blended
with the doublet \FeII~$\lambda 9956+9998$ on its blue side and with
\HeII~1.0124 $\mu$m or, in objects with strong iron emission,
\FeII~1.0132 $\mu$m and \FeII~1.0174 $\mu$m on its red side (see
Fig. \ref{PaAPaBPaD}). The strongest narrow emission line in the
near-IR, namely, \SIII~$\lambda 9531$ contaminates the peak and,
therefore, in particular the narrow component of the Pa$\epsilon$
emission line. In addition, in strong iron emitters, \FeII~$\lambda
9407$ and \FeII~$\lambda 9573$ blend in on the blue side and close to
the emission line center, respectively (see Fig. \ref{PaAPaBPaE}).

In Fig. \ref{PaBalFWHM} we also compare the widths of the Paschen and
Balmer emission lines. We chose to plot Pa$\beta$, since this line is
observed relatively unaffected by atmospheric absorption, versus the
two strongest Balmer emission lines, namely, H$\alpha$ and H$\beta$
(open triangles and squares, respectively). No significant difference
is seen between the emission line widths of H$\alpha$ and Pa$\beta$.
However, there is a trend for the widths of the H$\beta$ broad
components to be larger than those of the Pa$\beta$ broad
components. This is most likely due to the well-known existence of a
``red shelf'' in the H$\beta$ broad component \citep[e.g.,][]{DeR85,
Marz96}, which is believed to be formed, at least in part, by emission
from weak \FeII~multiplets and \HeI~$\lambda 4922, 5016$
\citep{Veron02, Koll01}. The H$\beta$ ``red shelf'' is observed in the
majority of our sources, and especially the three objects with the
largest width differences between the H$\beta$ and Pa$\beta$ broad
components, namely, Mrk 876, Mrk 590, and NGC 5548, clearly show this
feature.

\begin{figure*}
\centerline{ \includegraphics[scale=0.34]{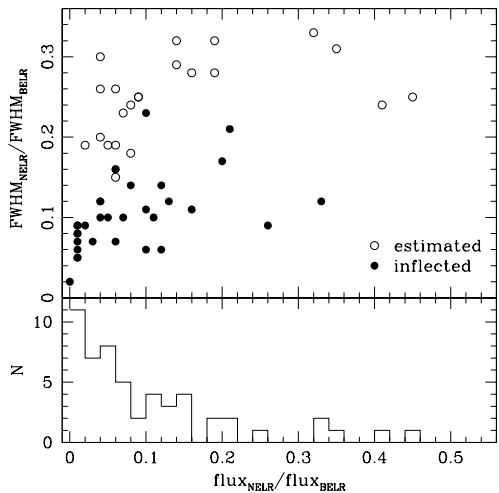} }
\caption{\label{NLRBLR}  Upper panel: The ratio between
the full width at half maximum (FWHM) of the narrow and broad
components versus their flux ratios for the cases where the narrow
component was inflected (filled circles) and estimated (open
circles). Included are the emission lines Pa$\beta$, H$\alpha$ and
H$\beta$. Lower panel: A histogram of the flux ratios between the
narrow and broad components plotted in the upper panel.}
\end{figure*}

This study is the first to systematically separate the narrow and
broad components for both a large number of emission lines and a
relatively large sample of type 1 AGN. Taking the Pa$\beta$,
H$\alpha$, and H$\beta$ emission lines as representative, we find that
in the majority of the cases the flux of the narrow component is less
than $\sim 10\%$ of the flux of the broad component
(Fig. \ref{NLRBLR}, lower panel). Whether the narrow component appears
inflected or not is, as expected, independent of the flux ratio
between the narrow and broad components and depends only on their
relative widths. From Fig. \ref{NLRBLR}, upper panel, we obtain that
the narrow component appears inflected when its width is below $\sim
20\%$ of that of the broad component.

\subsubsection{Dust Extinction}

Although the contribution of the narrow component to the total
emission line flux is small compared to that of the broad component
(see above), the physical conditions of the NELR are, contrary to
those of the BELR, relatively well understood
\citep[e.g.,][]{Osterbrock2}. Therefore, independent measurements of
the narrow component fluxes can yield important constraints on, e.g.,
the presence of dust within our line of sight which could also affect
the BELR.

\begin{figure*}
\centerline{ \includegraphics[scale=0.46]{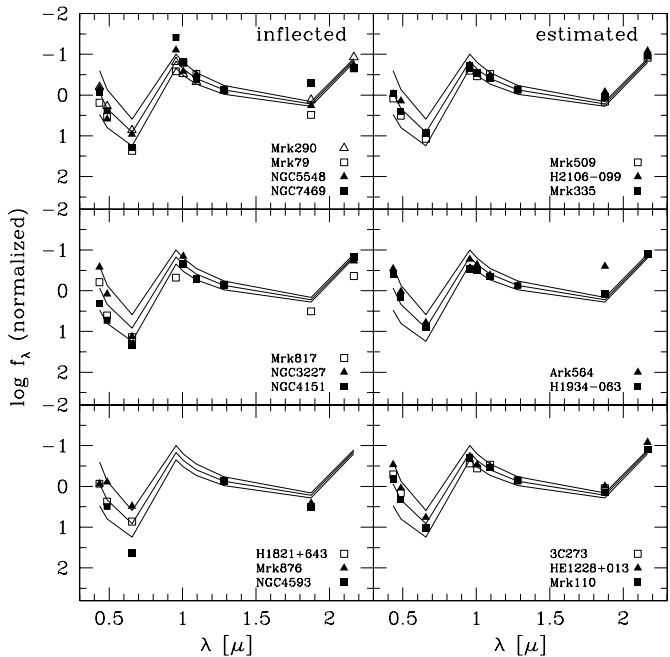} }
\caption{\label{NLRCaseB}  Narrow component emission line
fluxes compared to expectations from Case B recombination (for a
temperature of $T=15000$ K and an electron density of $n_e=10^4$
cm$^{-3}$). From bottom to top the dust extinction assumed was $A_V =
0, 1$ and 2 mag. The measurement points and the Case B solid lines
were normalized to the flux of Pa$\beta$.}
\end{figure*}

In Fig. \ref{NLRCaseB} we compare our measured narrow emission line
fluxes with the predictions from Case B recombination without dust
(lower solid line), and including a dust extinction of $A_{\rm V}=1$
and 2 mag (middle and upper solid lines, respectively). The values
expected from Case B recombination were calculated using the CLOUDY
photoionization simulation code \citep[last described by][]{Cloudy}
and assuming a temperature of $T=15000$ K and an electron density of
$n_e=10^4$ cm$^{-3}$. The $A_{\rm V}$ values were transformed into
$A_{\lambda}$ values using the analytical expression for the
interstellar extinction curve of \citet{Car89} and assuming a
parameter of $R_V = 3.1$. Fig. \ref{NLRCaseB} shows the emission line
fluxes normalized such that the flux of the Pa $\beta$ emission line
(the near-IR line measured for all sources) was unity. The Case B
curves were normalized such that the $A_{\rm V}=1$ mag curve at the
wavelength of Pa $\beta$ was unity. Since the measurement errors for
the Pa $\beta$ data point (roughly the size of the symbol) correspond
roughly to the separation between this curve and those for $A_{\rm
V}=0$ and 2 mag, this representation simply makes the Pa $\beta$ data
point (and in general the near-IR data points) compatible with Case B
recombination, irrespective of the amount of dust.

Fig. \ref{NLRCaseB} shows that the observed narrow emission line
ratios can be well reproduced by Case B recombination. The effects of
dust, however, are clearly observable only in the Balmer emission line
ratios, whereas they cannot be detected at a significant level using
the Paschen and Brackett emission lines alone. We then assessed the
extinction affecting the NELR in our sources based on which of the
plotted curves fitted the Balmer emission lines best. For the large
majority of our sources plotted in Fig. \ref{NLRCaseB} (12/18 objects)
we obtain that the dust extinction is of the order of $A_{\rm V} \sim
1$ mag. This average seems to be independent of the type of narrow
component (i.e., inflected or estimated). For three sources, namely,
Mrk 817, NGC 4151, and NGC 4593, the dust extinction is much below
this value, whereas for another three sources, namely, Mrk 876, Ark
564, and HE 1228$+$013, it appears to be as high as $A_{\rm V} \sim 2$
mag.

Significant reddening of the narrow emission lines of the order
obtained here has been observed before in type 1 AGN. \citet{Cohen83}
measured the reddening in the inflected narrow components of 13 type 1
AGN using two to four Balmer lines. Correcting his values for Galactic
extinction we obtain for his sample (which has 5 sources in common
with ours) an average of $A_{\rm V} \sim 1$ mag, similar to our
result. \citet{Ward87} estimated the NELR reddening of four objects,
of which two, namely, NGC 3227 and NGC 4151, are included in our
sample. These authors obtain on average a higher extinction of $A_{\rm
V} \sim 2$ mag. More recent studies, considering, however, only single
objects, obtain values similar to our average \citep[e.g.,][]{Cren01,
Cren02}.

Given this result, the relevant question for our studies is whether
the dust causing the observed extinction of the NELR affects also the
broad components. This will depend on the location of the dust, i.e.,
if it is mixed in with the gas producing the narrow emission lines
(``internal dust'') or if it is located outside the NELR, e.g., in the
host galaxy. Since the covering factor of the narrow emission line
clouds is assumed to be very small (only a few per cent), our line of
sight towards the BELR will not necessarily intercept these. However,
dust external to the NELR will act like a screen and affect also the
smallest scale components such as the BELR and the continuum emitted
by the accretion disk. In a future paper we plan to study in detail
the continuum spectral energy distributions of our sample,
supplementing our simultaneous optical and near-IR observations with
UV and X-ray archival observations. This will help determine if the
continuum source suffers from similarly strong extinction as the
NELR.

\subsection{The Helium Emission Lines}

\begin{figure*}
\centerline{
\includegraphics[clip=true,bb=1 545 590 710,scale=1.0]{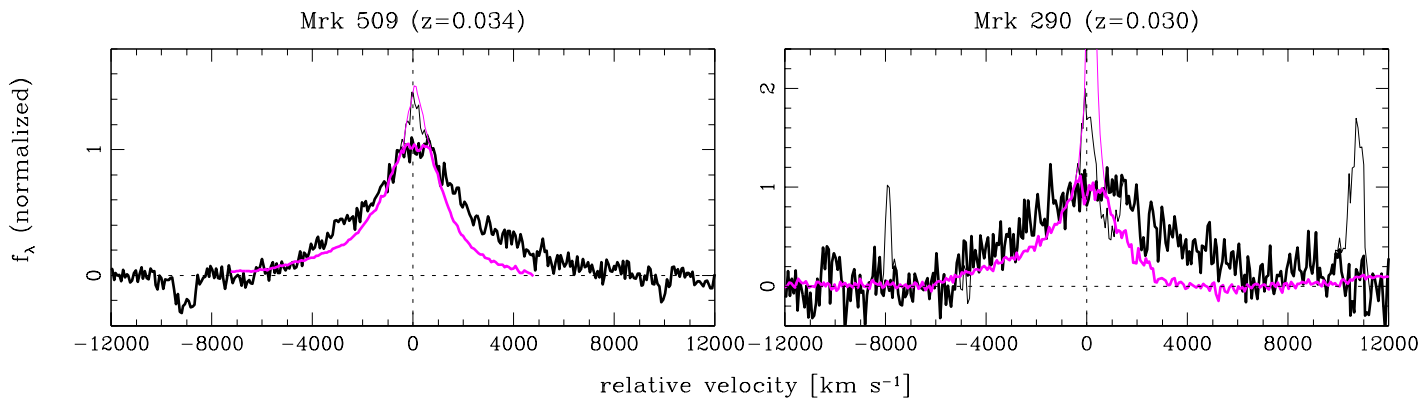}
}
\caption{\label{IRHeI}  Profiles of the broad emission
lines \HeI~$\lambda 5876$ (black) and \HeI~1.0830 $\mu$m (magenta) in
velocity space relative to the expected rest frame wavelength (thick
lines). The profiles have been continuum-subtracted and normalized to
the same peak intensity (of the broad component). The subtracted
blends and narrow components are also shown (thin lines).}
\end{figure*}

The helium transitions observed in the near-IR are by far less
numerous than the hydrogen emission lines, with only few lines clearly
detectable for both neutral and singly-ionized helium. We have
measured the strongest of the near-IR emission lines of neutral
helium, namely, \HeI~1.0830 $\mu$m, and also the strongest of the
optical emission lines, namely, \HeI~$\lambda 5876$. The \HeI~1.0830
$\mu$m emission line is blended with Pa $\gamma$ (and in strong iron
emitters also with \FeII~1.0863 $\mu$m), but, since it is a strong
line, it can be easily separated from this using our template profile
technique. The \HeI~$\lambda 5876$ emission line is affected by host
galaxy absorption, and we have corrected for it as described in
Section \ref{host}. Conclusive evidence for the presence of
\NaI~$\lambda 5896$ in emission is observed only in HE 1228$+$013. In
Fig. \ref{IRHeI} we compare the isolated profiles of the two
\HeI~emission lines and discuss these further below.

The near-IR (and also optical) \HeII~emission lines, and in particular
their broad components, are not only relatively weak, but suffer from
strong blending. Therefore, the extraction and measurement of the
profiles of these emission lines are best performed on mean and root
mean square (rms) spectra. In such spectra the constant or slowly
varying components, such as, e.g., the wings of a (blending) broad
emission line, do not appear, and the \HeII~emission line can be
reliably isolated \citep[e.g.,][]{Pet99}. In a future paper we will
present results from our near-IR reverberation mapping campaign and
will discuss in detail the profiles of the \HeII~emission lines.

\subsubsection{The Kinematics of the BELR}

The helium and hydrogen emission lines can be used together to study
the velocity structure of the BELR. Since emission lines corresponding
to higher ionization potentials are expected to be produced at
locations closer to the AGN's central ionizing source than emission
lines corresponding to lower ionization potentials \citep[e.g.,][and
references therein]{Pet93, Ost86}, helium is expected to emanate from
locations closer to the center than those responsible for the hydrogen
emission. Thus, comparison of the widths of the helium and hydrogen
broad emission lines can give important clues regarding the kinematics
of the emitting region.

\begin{figure*}
\centerline{ 
\includegraphics[scale=0.44]{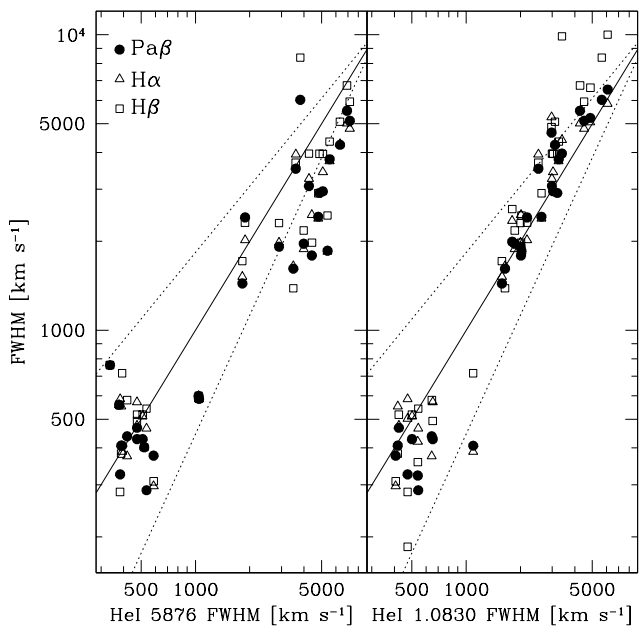} 
}
\caption{\label{HeIFWHM}  The full width at half maximum
(FWHM) of the broad and narrow {\sl inflected} components (upper right
and lower left symbols, respectively) of \HeI~$\lambda 5876$ and
\HeI~1.0830 $\mu$m vs. the hydrogen emission lines Pa$\beta$,
H$\alpha$, and H$\beta$ (filled circles, open triangles, and open
squares, respectively). Emission lines are considered to have similar
widths (solid line) if they lie within the error range given by the
spectral resolution (dotted lines).}
\end{figure*}

In Fig. \ref{HeIFWHM} we plot the widths of the two \HeI~emission
lines versus those of the hydrogen emission lines Pa$\beta$,
H$\alpha$, and H$\beta$, separately for their (inflected) narrow and
broad components. Fig. \ref{HeIFWHM}, left panel, shows that whereas
the narrow components of the helium and hydrogen emission lines have
similar widths, the broad component of \HeI~$\lambda 5876$ is
significantly wider than that of the hydrogen emission lines. E.g.,
the average FWHM ratio between the \HeI~$\lambda 5876$ and Pa$\beta$
emission lines is $\sim 1.5$. This result is similar to previous
findings \citep[e.g.,][]{Ost82, Shang07} and in accordance with the
current picture that the BELR is dominated by Keplerian motion in the
gravitational field of a central supermassive black hole
\citep[e.g.,][]{Pet99}.

On the other hand, the broad component of \HeI~1.0830 $\mu$m has
widths rather similar to those of the hydrogen emission lines
(Fig. \ref{HeIFWHM}, right panel). This is expected since \HeI~1.0830
$\mu$m is one of the most optically thick helium emission lines, and,
therefore, its emissivity should drop off significantly with
increasing incident photon flux. This then renders the \HeI~1.0830
$\mu$m emission line a less suitable probe of the BELR kinematics than
the \HeI~$\lambda 5876$ emission line.

\subsection{The Oxygen Emission Lines}

\begin{figure*}
\centerline{
\includegraphics[clip=true,bb=1 519 590 710,scale=1.0]{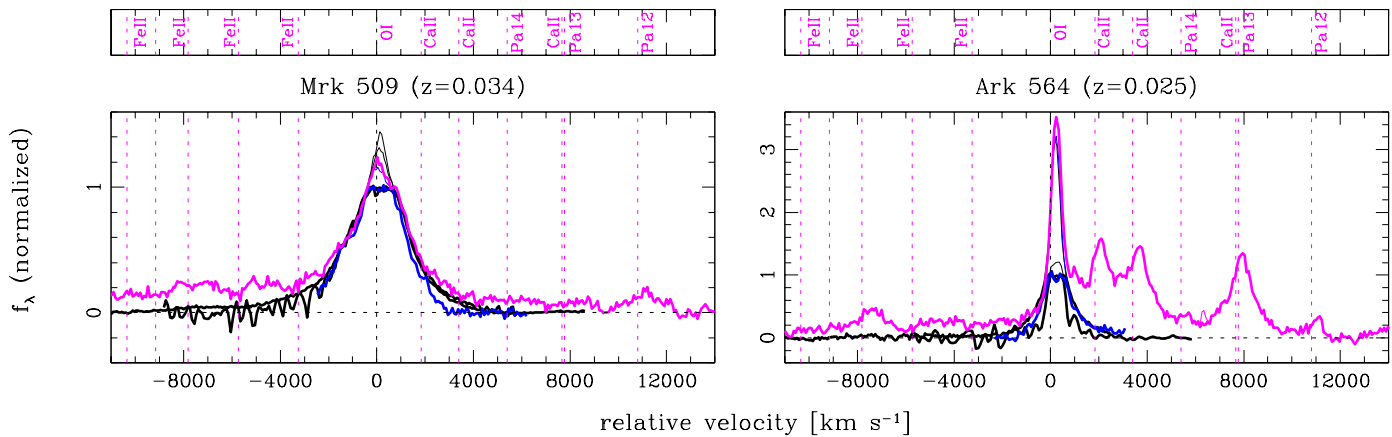}
}
\caption{\label{IROI}  IRTF SpeX profiles of the broad
emission lines \OI~$\lambda 8446$ (magenta) and \OI~1.1287 $\mu$m
(blue) compared to the profiles of Pa$\alpha$ and Pa$\beta$ (black) in
velocity space relative to the expected rest frame wavelength (thick
lines). The \OI~$\lambda 8446$ emission line blends with the
IR~\FeII~'hump', \CaII~triplet, Pa14, Pa13, and Pa12 (magenta vertical
dotted lines). The profiles have been continuum-subtracted and
normalized to the same peak intensity (of the broad component). The
subtracted narrow components and blends of narrow emission lines are
also shown (thin lines). [{\it The spectral plots for all sources are
available in colour in the electronic edition of the journal.}]}
\end{figure*}

We clearly observe in the near-IR spectra of all our sources two of
the strongest emission lines of neutral oxygen, namely, \OI~$\lambda
8446$ and \OI~1.1287 $\mu$m (Fig. \ref{IROI}). Contrary to the
hydrogen and helium emission lines discussed in the previous sections,
which result mainly from recombination processes, the observed oxygen
transitions are assumed to be the result of Ly$\beta$ fluorescence
\citep[e.g.,][]{Grandi80, Rudy89}. However, it is not clear if other
excitation processes can also contribute \citep{Rod02b}. In addition,
contrary to hydrogen and helium, neutral oxygen emission is believed
to be a pure broad-line phenomenon \citep[][]{Grandi80, Mor89}.

In the following we report for the first time the detection of narrow
components to \OI~emission lines and address the issue of the
\OI~excitation mechanism. Understanding this mechanism is essential
since \OI~emission lines are believed to be a suitable reddening
indicator for the BELR \citep{Netz79}.

\subsubsection{The Presence of Narrow Components}

\citet{Grandi80} was the first to observe the spectral region around
\OI~$\lambda 8446$ for a relatively large sample of Seyfert 1 galaxies
(16 objects) and for three Seyfert 2 galaxies.  Based on the lack of a
detection of \OI~$\lambda 8446$ in the Seyfert 2 galaxies and the lack
of a narrow component to \OI~in intermediate Seyfert galaxies (i.e.,
in Seyfert galaxies with otherwise inflected narrow components), such
as, e.g., NGC 5548 and NGC 4151, he concluded that \OI~is emitted in
the BELR only.

Like \citet{Grandi80}, we do not observe narrow components of the
\OI~emission lines in sources which have hydrogen emission lines with
inflected narrow components. However, in all other sources, the
\OI~profiles are similar to those of the {\sl total} hydrogen profiles
(Fig. \ref{IROI}). This suggests that \OI~emission can be produced in
the NELR. In support of this interpretation are also the findings of
\citet{Mor85}, who found \OI~emission in two Seyfert 2 galaxies,
namely, NGC 5506 and NGC 7314.

\subsubsection{The Excitation Mechanism}

There are four excitation mechanisms which can lead to the production
of \OI~$\lambda 8446$ emission: recombination, collisional excitation,
continuum fluorescence, and Ly$\beta$ fluorescence
\citep[e.g.,][]{Grandi75, Grandi80}. The \OI~1.1287 $\mu$m emission
line, on the other hand, is believed to be the result of Ly$\beta$
fluorescence only.

In order to determine the principal excitation mechanism for
\OI~$\lambda 8446$ in our sources, we start by comparing the {\sl
photon} fluxes of the two oxygen emission lines. If Ly$\beta$
fluorescence dominates, every \OI~$\lambda 8446$ photon is produced
via an \OI~1.1287 $\mu$m photon, and, therefore, we expect the photon
flux ratio between the two oxygen emission lines to be unity. If,
however, other processes also contribute, the \OI~$\lambda 8446$
emission will be enhanced relative to that of \OI~1.1287 $\mu$m.

\begin{figure*}
\centerline{
\includegraphics[scale=0.42]{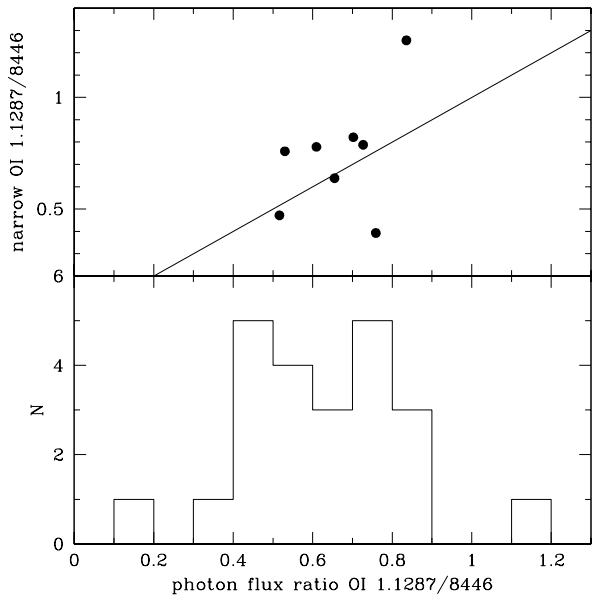}
}
\caption{\label{OIratios}  Lower panel: The distribution
of the photon flux ratios between the \OI~1.1287 $\mu$ and
\OI~$\lambda 8446$ emission lines for the broad components. Upper
panel: The photon flux ratios between the broad components of the
\OI~1.1287 $\mu$ and \OI~$\lambda 8446$ emission lines vs. those of
the narrow components.}
\end{figure*}

Fig. \ref{OIratios}, lower panel, shows the distribution of the photon
flux ratios between the broad components of the \OI~1.1287 $\mu$m and
\OI~$\lambda 8446$ emission lines. In all sources, with the exception
of Mrk 590, this ratio is below unity, indicating that processes other
than Ly$\beta$ fluorescence contribute to the broad \OI~$\lambda 8446$
emission. For Mrk 590 we obtain a photon flux ratio of 1.13. Ly$\beta$
fluorescence contributes in our sources between $\sim$20-85\% of the
broad \OI~$\lambda 8446$ emission, with 14/23 sources having a
contribution from this process of $\ga$60\%. In only three sources
does this process not dominate, namely, in PDS 456 ($\sim$20\%), Ark
120 ($\sim$40\%), and NGC 5548 ($\sim$40\%), and in the remaining six
sources, namely, Mrk 876, Mrk 110, Mrk 509, Mrk 817, Mrk 290, and NGC
4151, Ly$\beta$ fluorescence accounts for only half of the broad
\OI~$\lambda 8446$ flux.

In Fig. \ref{OIratios}, upper panel, we compare for the eight sources
with detected narrow \OI~components the photon flux ratios between the
\OI~1.1287 $\mu$m and \OI~$\lambda 8446$ emission lines in the BELR
and NELR. Three sources, namely, Mrk 509, H 2106$-$099, and Ark 564,
have in both emission line regions similar contributions from
Ly$\beta$ fluorescence of $\sim$50\%, $\sim$65\%, and $\sim$75\%,
respectively. In four other sources, namely, HE 1228$+$013, Mrk 110,
Mrk 335, and H 1934$-$063, the Ly$\beta$ fluorescence contribution is
higher in the NELR than in the BELR, with the largest differences
found for Mrk 110 ($\sim$75\% in the NELR vs. $\sim$50\% in the BELR)
and HE 1228$+$013 (a larger than unity photon flux ratio of 1.26 in
the NELR vs. $\sim$85\% in the BELR). For 3C 273 we find a smaller
Ly$\beta$ fluorescence contribution in the NELR than in the BELR
($\sim$40\% vs. $\sim$75\%).

Since Ly$\beta$ fluorescence cannot account for all of the
\OI~$\lambda 8446$ emission line flux, neither in the BELR nor in the
NELR, which of the other three possible excitation mechanisms could
also contribute? This can be best assessed by the presence or lack of
other \OI~emission lines. Continuum fluorescence implies the presence
of lines such as, e.g., \OI~1.3165 $\mu$m, which is the strongest line
produced by this process. Recombination and collisional excitation
produce, e.g., \OI~$\lambda 7774$ (the quintet counterpart to
\OI~$\lambda 8446$), with ratios of \OI~$\lambda 7774/\lambda
8446\approx1.1$ (taking the transition probabilities given in the
Atomic Line List v2.04) and $\approx 0.3$ \citep{Grandi80},
respectively.

\begin{figure*}
\centerline{
\includegraphics[clip=true,bb=1 295 590 718,scale=1.0]{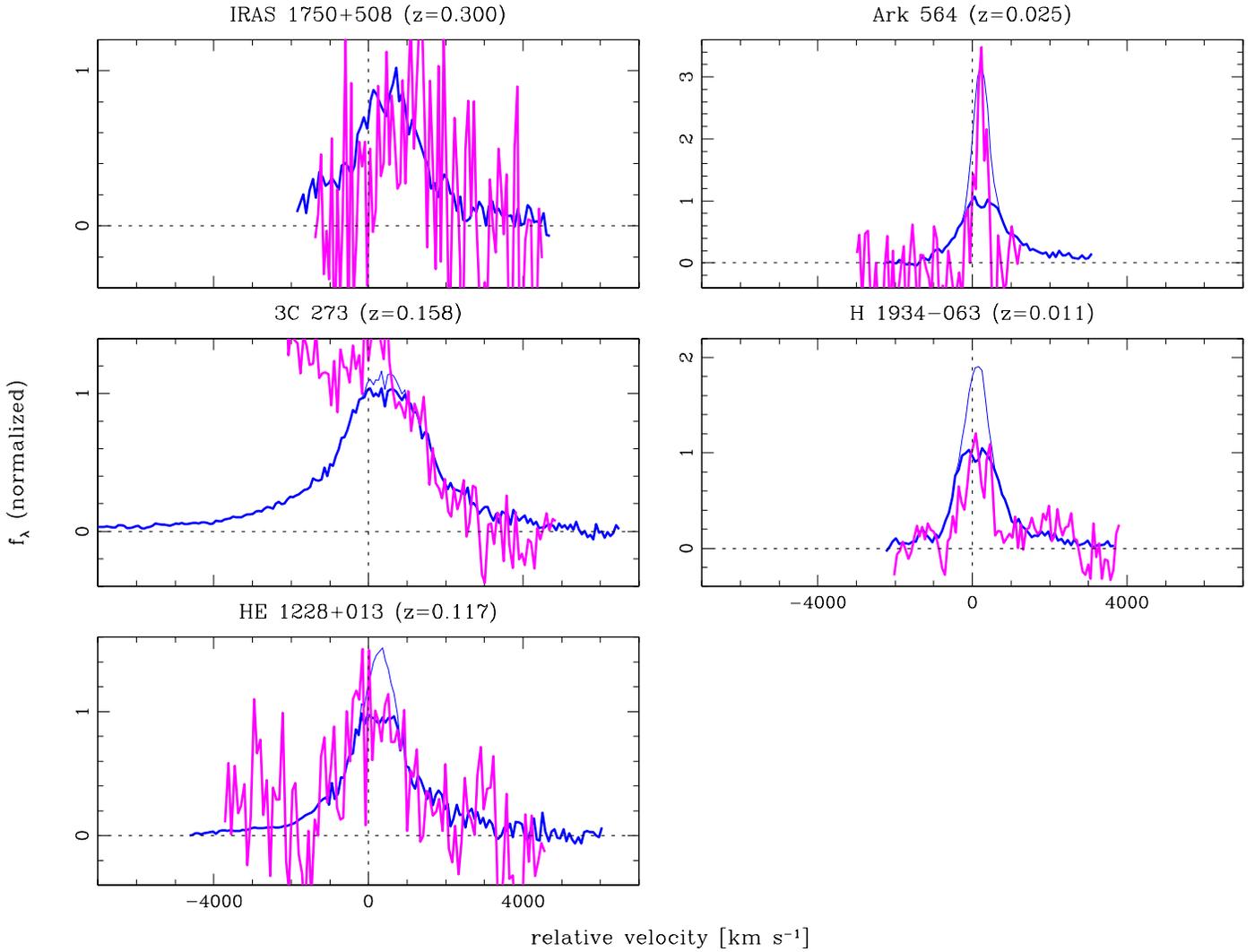}
}
\caption{\label{OI13168}  IRTF SpeX profiles of the broad
emission lines \OI~1.3165 $\mu$m (magenta) and \OI~1.1287 $\mu$m
(blue) in velocity space relative to the expected rest frame
wavelength (thick lines). The profiles have been continuum-subtracted
and normalized to the same peak intensity (of the broad
component). The subtracted narrow components of \OI~1.1287 $\mu$m are
also shown (blue thin lines).}
\end{figure*}

The near-IR spectra of the seven sources in our sample with the
highest redshifts cover the spectral region around \OI~$\lambda 7774$
and this emission line is clearly detected in all of them (see
Fig. \ref{OI7776}). \OI~$\lambda 7774$ is blended with \FeII~$\lambda
7712$ and an unidentified feature which appears to be itself a blend
of emission lines at $\sim 7875$~\AA~and $\sim 7896$~\AA~(see Section
\ref{noID}). Therefore, we have measured its flux using the profile of
\OI~1.1287 $\mu$m, similar to the procedure adopted for \OI~$\lambda
8446$ (see Section \ref{blends} for details). The \OI~1.3165 $\mu$m
emission line is convincingly detected in only 5/15 sources with
spectral coverage in this region (see Fig. \ref{OI13168}). In 3C 273
this line is blended with the red wing of Pa$\beta$ and we used the
\OI~1.1287 $\mu$m profile to measure it. 

In Table \ref{oxygenex} we list for a total of nine sources the fluxes
for the broad components of \OI~1.3165 $\mu$m and \OI~$\lambda 7774$
(columns (4) and (6), respectively). A narrow component to the
\OI~$\lambda 7774$ emission line was detected in 3C 273, but not in HE
1228$+$013. Similarly, \OI~1.3165 $\mu$m narrow components were
detected in 3C 273 and Ark 564, but not in HE 1228$+$013 and H
1934$-$063. Of the nine sources listed in Table \ref{oxygenex}, seven
have their broad \OI~$\lambda 8446$ emission dominated by Ly$\beta$
fluorescence (see column (3)). However, for PDS 456 this process is
important only at the $\sim$20\% level and for Mrk 876 it accounts for
only about half of the broad \OI~$\lambda 8446$ flux.

For 3/9 sources, namely, IRAS 1750$+$508, 3C 273, and HE 1228$+$013,
both \OI~1.3165 $\mu$m and \OI~$\lambda 7774$ were detected, Based on
their photon flux ratios relative to \OI~$\lambda 8446$, continuum
fluorescence seems to contribute only a few per cent and recombination
seems to account for the remaining $\sim 10-20$\% of the broad
\OI~$\lambda 8446$ emission line flux. However, for 3C 273 the
situation in its NELR is different from that in its BELR. As mentioned
above, Ly$\beta$ fluorescence accounts for only $\sim$40\% of its
narrow \OI~$\lambda 8446$ emission, and, whereas continuum
fluorescence contributes only a few per cent, collisional excitation
seems to be the dominant process.  For the narrow components we obtain
a photon flux ratio of \OI~$\lambda 7774/\lambda 8446 \sim 0.21$,
after subtracting the fluorescence contribution off the \OI~$\lambda
8446$ flux.

For another 3/9 sources, namely, H 1821$+$643, Mrk 876, and PDS 456,
the spectrum covered the location of both \OI~1.3165 $\mu$m and
\OI~$\lambda 7774$, but only \OI~$\lambda 7774$ was detected. In H
1821$+$643 and Mrk 876 recombination seems to account for the
remaining $\sim 20-50$\% of the broad \OI~$\lambda 8446$ emission line
flux, but PDS 456 has a photon flux ratio of \OI~$\lambda 7774/\lambda
8446 \sim 0.24$ (after subtracting the fluorescence contribution off
the \OI~$\lambda 8448$ flux), similar to the value expected for
collisional excitation.

In PG 0844$+$349, the spectral region around \OI~1.3165 $\mu$m was not
covered, but the photon flux ratio between \OI~$\lambda 7774$ and
\OI~$\lambda 8446$ indicates that recombination accounts for the
remaining broad \OI~$\lambda 8446$ flux. On the other hand, in Ark 564
and H 1934$-$063, the spectral region around \OI~$\lambda 7774$ was
not covered, but the photon flux ratio between \OI~1.3165 $\mu$m and
\OI~$\lambda 8446$ indicates that continuum fluorescence is not as
important as recombination or collisional excitation.

In summary, Ly$\beta$ fluorescence is the dominant production
mechanism of broad \OI~$\lambda 8446$ in just over half of our sources
(14/23 objects). In roughly a third of our sources (6/23 objects)
Ly$\beta$ fluorescence accounts for about half of the broad
\OI~$\lambda 8446$ flux, and in only a few sources it is not the
dominant process. In one of these few sources, namely, PDS 456,
collisional excitation dominates instead. For six sources we determine
that recombination (and not collisional excitation) accounts for the
remaining broad \OI~$\lambda 8446$ emission line flux. Continuum
fluorescence is present, but does not seem to be important.

\citet{Rod02b} were the first to present evidence that Ly$\beta$
fluorescence is not the only process contributing to the \OI~$\lambda
8446$ emission in AGN. We confirm their result using a sample more
than three times as large. Similar to our findings, \citet{Rod02b}
concluded that continuum fluorescence does not contribute
significantly, but since they had observations of \OI~$\lambda 7774$
for only one source they could not determine which of the two,
recombination or collisional excitation, was in general the other
important process. Their only source with available \OI~$\lambda 7774$
was H 1934$-$063, which is also part of our sample. This line was
covered by their optical spectrum which extended further into the red
than ours. For this source they obtain a photon flux ratio of
\OI~$\lambda 7774/\lambda 8446 \sim 0.2$ (after subtracting the
fluorescence contribution off the \OI~$\lambda 8446$ flux), suggesting
that collisional excitation accounts for the remaining $\sim$30\% of
the \OI~$\lambda 8446$ flux.

Recently, \citet{Mat07} presented \OI~$\lambda 8446$ and \OI~1.1287
$\mu$m emission line measurements for six quasars. They also concluded
that \OI~$\lambda 8446$ emission is not produced by Ly$\beta$
fluorescence alone and identified collisional excitation as the other
important process. However, this result was obtained based on
comparisons of their observed \OI~emission line ratios with
predictions from photoionization models, which did not include the
contribution from recombination. Thus they could only distinguish
between collisional excitation and continuum fluorescence. In
addition, we note that for the one source in common with our sample,
namely, 3C 273, they measured a similar \OI~$\lambda 8446$ flux, but a
\OI~1.1287 $\mu$m flux a factor of $\sim 2$ lower than our value.

\subsection{The Iron Emission Lines} \label{iron}

We have detected in our near-IR spectra numerous emission lines from
the rich spectrum of singly-ionized iron. Contrary to the
\FeII~emission lines seen in the UV and optical spectra of AGN, which
often blend together and at places form a pseudo-continuum, several
near-IR \FeII~emission lines can be resolved and studied individually.
This is important for at least three reasons: (i) the widths of the
resolved near-IR \FeII~emission lines can be used to broaden suitable
UV and optical templates, which leads to a more reliable iron
subtraction and thus measurement of the properties of contaminated
features. We have used this method successfully in Section
\ref{measurements}.; (ii) the fluxes of the resolved near-IR
\FeII~emission lines can be used to test theoretical iron emission
models. We do this in Section \ref{iron1}.; and (iii) the profiles of
the resolved near-IR \FeII~emission lines can be compared to those of
emission lines of other elements to help answer the long-standing
question of where the iron in AGN is produced and excited (see Section
\ref{iron2}).

\subsubsection{Observed vs. Predicted Line Strengths} \label{iron1}

Theoretical iron flux templates have been developed in recent years in
order to alleviate the problems inherent to empirical templates. Such
problems are, e.g., the assumption that the source used to construct
the template is typical of the AGN population, and the complications
introduced by the severe blending of broadened AGN emission lines
which makes the isolation of an iron template a difficult
task. Furthermore, theoretical iron templates have the advantage that
model parameters can be adjusted to fit observations, thus allowing
one to constrain the excitation mechanism producing the
\FeII~spectrum.

The most elaborate theoretical predictions were presented by
\citet{Sig03}. Given the relatively large size of our sample and the
quality of our data we can now put these to test. For this purpose we
have measured in our spectra the fluxes of six individual near-IR
\FeII~emission lines and of three blends. In addition, we have
measured the flux of the optical \FeII~'hump' in the range 4440-4700
\AA~as detailed in Section \ref{measurements} and of the near-IR
\FeII~'hump' in the range 8150-8365 \AA. Our results are listed in
Table \ref{iriron}. Two sources, namely, H 1821$+$643 and NGC 5548,
were omitted from this table since their \FeII~emission was weak and
only the optical and near-IR emission 'humps' could be convincingly
detected. In Table \ref{iriron} we also give the measured fluxes
relative to that of the \FeII~1.0491$+$1.0502 $\mu$m blend. We chose
this blend since it is relatively strong and observed unaffected by
emission from other elements.

In Table \ref{Feratios} we summarize for the six iron emission lines
with the largest statistics, namely, the two \FeII~'humps,
\FeII~$\lambda 9407$, \FeII~$\lambda 9573$, \FeII~$\lambda 9956+9998$
and \FeII~1.0174 $\mu$m, the average flux ratios relative to
\FeII~1.0491$+$1.0502 $\mu$m and also relative to \FeII~1.1126
$\mu$m. Although not observed in all our sources, we have considered
as a consistency check also ratios relative to the \FeII~1.1126 $\mu$m
emission line, since, contrary to \FeII~1.0491$+$1.0502 $\mu$m, it is
not a blend and similarly to this observed relatively free of
contaminating emission. We list in Table \ref{Feratios} averages for
the whole sample and also considering only those sources with the most
reliable \FeII~profiles for which all three of the emission lines
\FeII~$\lambda 9956+9998$, \FeII~1.0491$+$1.0502 $\mu$m, and
\FeII~1.1126 $\mu$m were detected (11/21 objects; see also
Fig. \ref{IRFeII}).

The observed means of all considered flux ratios relative to
\FeII~1.0491$+$1.0502 $\mu$m are significantly ($\ge 2 \sigma$)
different from the theoretical value expected by \citet{Sig03} for
their model A (see their Table 3). The only exception is
\FeII~$\lambda 9407$. The most significant deviations are observed for
the optical and near-IR \FeII~'humps' and for the \FeII~1.0174 $\mu$m
emission line. On the other hand, all considered flux ratios relative
to \FeII~1.1126 $\mu$m are consistent with theoretical
predictions. This is also the case for the flux ratio between the
optical and near-IR \FeII~'humps'.

The mean ratios of the optical and near-IR \FeII~'humps' relative to
\FeII~1.0491$+$1.0502 $\mu$m for the whole sample are larger than the
expected value by a factor of $\sim 2$ (significant at the $6 \sigma$
and $5 \sigma$ level, respectively). However, considering only the
sources with the most reliable \FeII~profiles the significance of the
difference between observations and predictions reduces to 2 $\sigma$.

The four near-IR \FeII~emission lines $\lambda 9956+9998$,
1.0491$+$1.0502 $\mu$m, 1.0174 $\mu$m, and 1.1126 $\mu$m descend from
a common upper multiplet and are collectively termed the ``1 $\mu$m
\FeII~lines'' \citep[e.g.,][]{Rudy00}. Both the \FeII~$\lambda
9956+9998$ and \FeII~1.1126 $\mu$m emission line fluxes are observed
versus that of \FeII~1.0491$+$1.0502 $\mu$m to be larger than
predicted by a factor of $\sim 2$ (at the $3 \sigma$ and $4 \sigma$
level, respectively), irrespective of the sample considered. The most
significant discrepancy between theory and observations is found for
the \FeII~1.0174 $\mu$m emission line. Its flux ratio relative to
\FeII~1.0491$+$1.0502 $\mu$m is on average a factor of $\sim 5$ larger
than predicted, significant at the $11 \sigma$ level. A similar result
is obtained if only the sources with the most reliable \FeII~profiles
are considered. The flux of this line relative to that of \FeII~1.1126
$\mu$m is larger than the predicted value by a smaller factor of $\sim
2$, a result which becomes significant only if the most reliable
measurements are considered.

In summary, our observations are consistent with the predictions of
\citet{Sig03} with two exceptions. The \FeII~1.0491$+$1.0502 $\mu$m
emission line flux is either overpredicted by theory or underestimated
by our measurements by a factor of $\sim 2$, and by a similar factor
the \FeII~1.0174 $\mu$m emission line flux is either underpredicted by
theory or overestimated by our measurements.

Considering in particular the sources with the most reliable
\FeII~profiles, it is unlikely that we have underestimated the flux of
\FeII~1.0491$+$1.0502 $\mu$m by as much as a factor of $\sim 2$, given
that this feature is observed relatively uncontaminated. On the other
hand, the measurement of the \FeII~1.0174 $\mu$m emission line is
complicated by its proximity to the rather strong \HeII~1.0124 $\mu$m
emission line. However, in the five sources with the most reliable
\FeII~profiles considered in Table \ref{Feratios}, namely, Mrk 110, H
2106$-$099, Mrk 335, Ark 564, and H 1934$-$063, the two profiles are
rather well separated (see also Fig. \ref{PaAPaBPaD}). Therefore, it
seems more likely that whereas the flux of \FeII~1.0491$+$1.0502
$\mu$m is overpredicted by theory, the flux of \FeII~1.0174 $\mu$m is
underpredicted, and this by a similar factor of $\sim 2$. This is
possible, if, e.g., the oscillator strengths used by \citet{Sig03} for
these two lines were in error. The \FeII~1.0491$+$1.0502 $\mu$m blend
is dominated by emission from the \FeII~1.0502 $\mu$m line, and this
line and \FeII~1.0174 $\mu$m originate from the same upper
level. Therefore, the population of this level seems to have been
calculated correctly, but not the depopulation
statistics. Alternatively, their computations might have to take into
account the importance of collisional transitions among highly excited
levels populated by Ly$\alpha$ fluorescence, particularly
pseudometastable levels, as suggested by \citet{Bau04}.

Near-IR \FeII~emission line measurements have been published so far
for only six AGN. \citet{Rudy00} and \citet{Rudy01} find for the
sources I Zw 1 and Mrk 478 values of $-$0.31 and $-$0.37,
respectively, for the logarithmic ratio between the \FeII~1.1126
$\mu$m and \FeII~1.0502 $\mu$m emission lines, similar to our mean,
but they do not measure the \FeII~1.0174 $\mu$m emission line
flux. However, they call attention to the finding of \citet{Rudy91}
who measured in the emission-line star Lk H$\alpha$ 101 the
\FeII~1.0174 $\mu$m flux to be stronger than predicted. The sample of
\citet{Rod02a} has 3/4 objects in common with our sample, namely, Mrk
335, Ark 564, and H 1934$-$063. For these we calculated based on the
values given in their Table 2 averages of $-$0.47$\pm$0.04,
$-$0.40$\pm$0.06, and $-$0.06$\pm$0.11 for the ratios log \FeII~1.0174
$\mu$m/1.0502 $\mu$m, log \FeII~1.1126 $\mu$m/1.0502 $\mu$m, and log
\FeII~1.0174 $\mu$m/1.1126 $\mu$m, respectively, which are similar to
our results.

\subsubsection{The Iron Emitting Region} \label{iron2}

Understanding where and how \FeII~emission is produced in AGN is
important for at least two reasons. Firstly, since the \FeII~atom has
a complex structure, it emits through a large number of multiplets,
thus acting as one of the main coolants of the BELR
\citep{Wills85}. Secondly, studies of \FeII~abundance changes in AGN
over cosmic time can give important clues regarding the chemical
evolution of the Universe \citep{Ham99}.

The excitation mechanisms believed to produce \FeII~most efficiently
are Ly$\alpha$ fluorescence and collisional excitation. In particular,
Ly$\alpha$ fluorescence is the process invoked for the production of
the near-IR \FeII~emission lines. In the previous subsection we
concluded that the theoretical predictions of \citet{Sig03} are
generally consistent with our observations, which suggests that
Ly$\alpha$ fluorescence is indeed the dominant process. In this case
then we also expect the profiles of the near-IR \FeII~emission lines
to be similar to those of hydrogen.

\begin{figure*}
\centerline{
\includegraphics[clip=true,bb=1 545 590 710,scale=1.0]{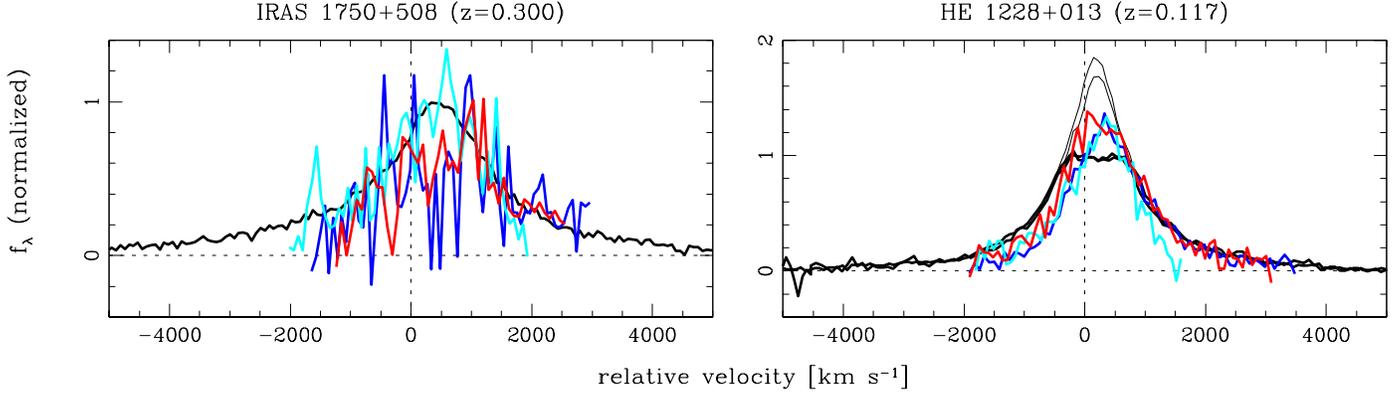}
}
\caption{\label{IRFeII}  IRTF SpeX profiles of the iron
emission lines \FeII~$9956+9998$ (cyan), \FeII~1.0491$+$1.0502 $\mu$m
(blue), and \FeII~1.1126 $\mu$m (red) compared to the profiles of
Pa$\alpha$ and Pa$\beta$ (black) in velocity space relative to the
expected rest frame wavelength (thick lines). The profiles have been
continuum-subtracted and normalized to the same peak intensity (of the
broad component). The subtracted narrow components of the Paschen
lines are also shown (thin lines). [{\it The spectral plots for all
sources are available in colour in the electronic edition of the
journal.}]}
\end{figure*}

In Fig. \ref{IRFeII} we compare the isolated profiles of the three
\FeII~emission lines $\lambda 9956+9998$ (cyan), 1.0491$+$1.0502
$\mu$m (blue), and 1.1126 $\mu$m (red) to those of the hydrogen
emission lines Pa$\alpha$ and Pa$\beta$ (black). About half of our
sources are relatively weak iron emitters and, therefore, their
\FeII~profiles are rather noisy. However, for the strongest iron
emitters Fig. \ref{IRFeII} clearly shows that the \FeII~and hydrogen
emitting regions are, as expected, cospatial. This result is similar
to that obtained by \citet{Bor92} based on comparisons of measured
H$\beta$ widths to modeled \FeII~widths. Therefore, in the absence of
near-IR spectra from which to measure the widths of the resolved
\FeII~emission lines, one can use the widths of available hydrogen
emission lines instead in order to broaden suitable \FeII~templates,
as is already common practice \citep[e.g.,][]{Marz96}.

\subsubsection{Unidentified Emission Lines} \label{noID}

For six features we could not find a suitable identification using the
\FeII~emission line list of \citet{Sig03}. However, comparisons of
their profiles to those of the two near-IR iron emission lines
\FeII~1.0491$+$1.0502 $\mu$m and \FeII~1.1126 $\mu$m, which are
relatively strong and observed unblended, show that they are most
likely so far unidentified iron emission lines. This interpretation is
strengthened by the fact that the unidentified features are
convincingly detected only in those eight sources that are amongst the
strongest iron emitters in our sample, namely, PDS 456, 3C 273, HE
1228$+$013, PG 0844$+$349, Mrk 817, Ark 564, Mrk 79, and H 1934$-$063.

In Table \ref{unidenttab} we list the rest frame wavelengths and
observed fluxes of these unidentified near-IR emission lines. The
given values, however, should be taken only as
approximations. Although we measured the rest wavelength in that
source where the feature appeared the narrowest, emission lines of AGN
are often red- or blue-shifted relative to the system velocity. The
measured flux, on the other hand, is overestimated, if the
unidentified feature is in fact a blend of several emission lines.

\begin{figure*}
\centerline{
\includegraphics[clip=true,bb=1 519 590 710,scale=1.0]{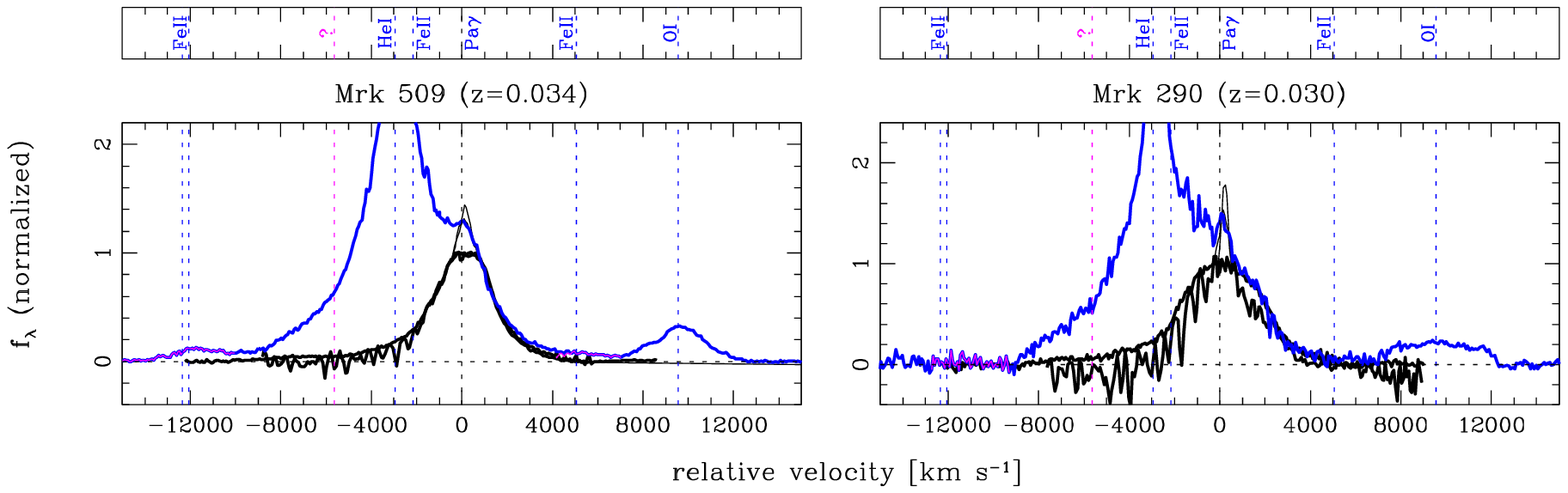}
}
\caption{\label{PaAPaBPaC}  IRTF SpeX profiles of the
emission line blend Pa$\gamma$, \HeI~1.0830 $\mu$m, \OI~1.1287 $\mu$m,
\FeII~1.0502 $\mu$m, \FeII~1.0863 $\mu$m, and \FeII~1.1126 $\mu$m
(blue, thick lines) compared to the profiles of Pa$\alpha$ and
Pa$\beta$ (black). Profiles are shown in velocity space relative to
the expected rest frame wavelength indicated by the vertical dotted
lines (black for the Paschen lines, blue for \HeI, \OI~and \FeII,
magenta for the unidentified feature). The profiles have been
continuum-subtracted and normalized to the same peak intensity (of the
broad component). The isolated \FeII~profiles (magenta) are also
shown. [{\it The spectral plots for all sources are available in
colour in the electronic edition of the journal.}]}
\end{figure*}

\begin{figure*}
\centerline{
\includegraphics[clip=true,bb=1 170 590 718,scale=1.0]{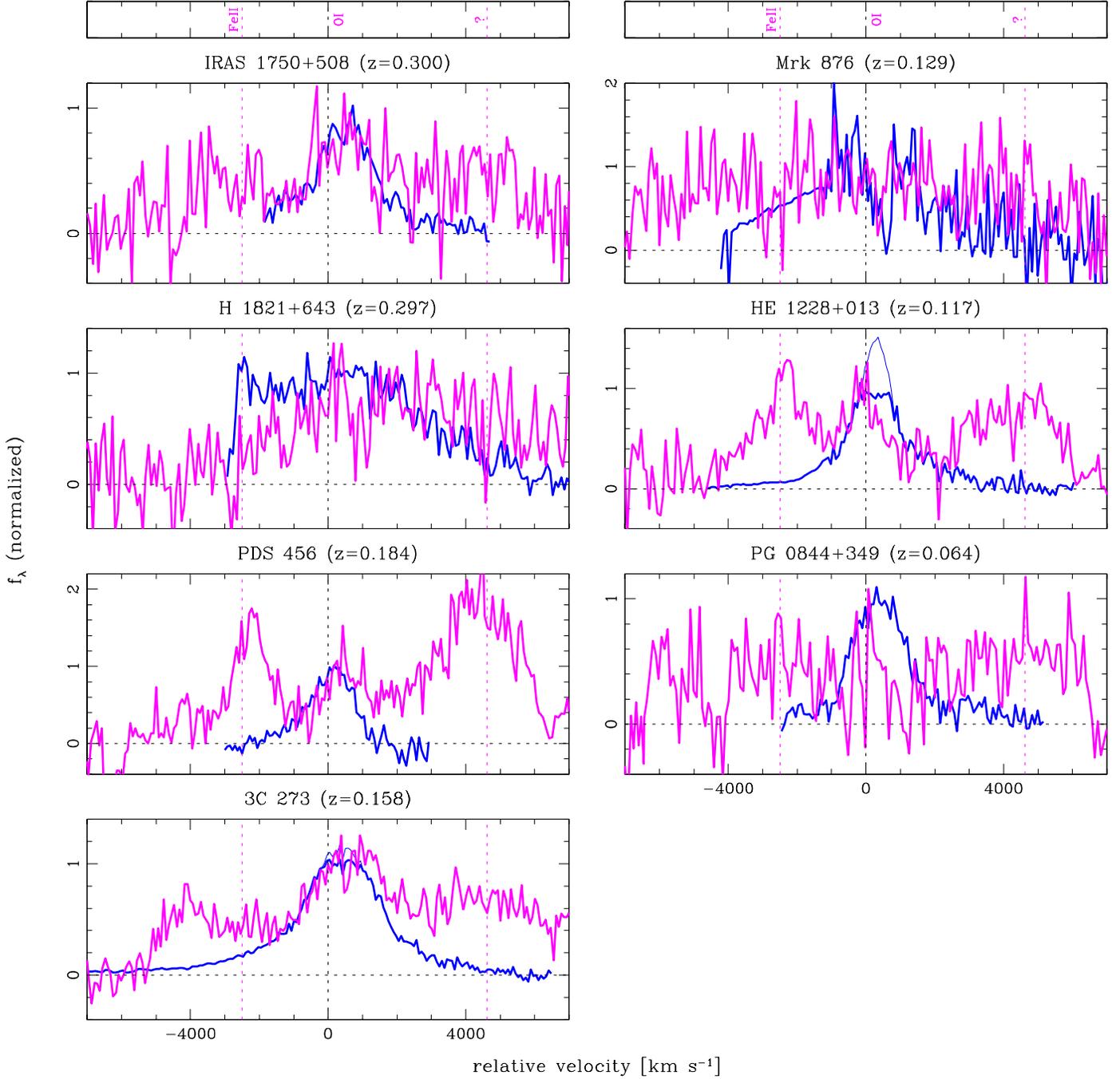}
}
\caption{\label{OI7776}  IRTF SpeX profiles of the broad
emission lines \OI~$\lambda 7774$ (magenta) and \OI~1.1287 $\mu$m
(blue) in velocity space relative to the expected rest frame
wavelength (thick lines). The \OI~$\lambda 7774$ emission line blends
with \FeII~$\lambda 7712$ and an unidentified feature (magenta
vertical dotted lines). The profiles have been continuum-subtracted
and normalized to the same peak intensity (of the broad
component). The narrow components of \OI~1.1287 $\mu$m are also shown
(blue thin lines).}
\end{figure*}

\begin{figure*}
\centerline{
\includegraphics[scale=0.46]{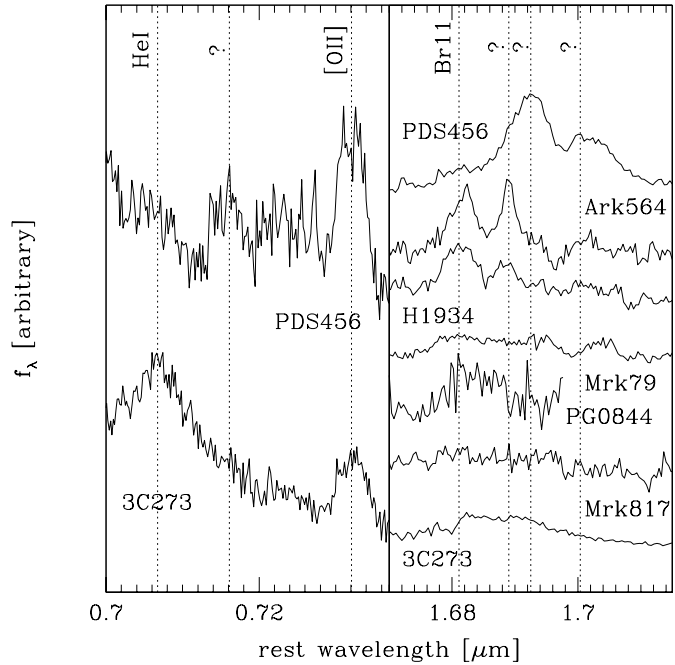}
}
\caption{\label{unidentfig}  IRTF SpeX profiles of the
unidentified emission features at 7161~\AA~(left panel) and at
16890~\AA, 16925~\AA, and 17004~\AA~(right panel).}
\end{figure*}

In Fig. \ref{unidentfig}, left panel, we show the feature at
7161~\AA. It is seen in PDS 456, but not in 3C 273, which is the only
other of the above listed sources for which this spectral range is
covered. The blends $\lambda 7875+7896$ and $\lambda 10703+10736$ are
labeled in Figs. \ref{OI7776} and \ref{PaAPaBPaC}, respectively. In
Fig. \ref{unidentfig}, right panel, we show the three unidentified
features at 16890~\AA, 16925~\AA, and 17004~\AA, which are most
clearly observed in PDS 456, Ark 564, H 1934$-$063, and Mrk 79. These
features are not convincingly detected in 3C 273, and the spectrum of
HE 1228$+$013 does not cover them. We note that the features at
16925~\AA~and 17004~\AA~in PDS 456 were observed also by
\citet{Sim99}.

\acknowledgments 

We thank Perry Berlind and Mike Calkins for the FAST observations and
Susan Tokarz for the reduction of these data. We thank Richard Binzel
for allowing us to use his IRTF remote observing set-up at
MIT. H.L. thanks St. John's College, Oxford, for its hospitality
during the last months of this work. H.L. acknowledges financial
support from the Deutsche Akademie der Naturforscher Leopoldina grant
number BMBF-LPD 9901/8-99. B.M.P. and M.C.B. are grateful for support
by the National Science Foundation through grants AST-0205964 and
AST-0604066 to the Ohio State University. M.C.B. is supported by a
National Science Foundation Graduate Fellowship. M.K. is a member of
the Chandra X-ray Center, which is operated by the Smithsonian
Astrophysical Observatory under NASA Contract NAS8-03060. This
research has made use of the NASA/IPAC Extragalactic Database (NED),
which is operated by the Jet Propulsion Laboratory, California
Institute of Technology, under contract with the National Aeronautics
Space Administration.

\bibliography{references}

\clearpage

\begin{deluxetable}{lcccccl}
\tabletypesize{\small} 
\tablecaption{
\label{general} 
General Properties of the Observed Sample} 
\tablewidth{0pt} 
\tablehead{ 
Object Name & R.A.(J2000) & Decl.(J2000) & z & J & V & class \\ 
&&&& [mag] & [mag] & \\
(1) & (2) & (3) & (4) & (5) & (6) & (7)
}
\startdata
Mrk 335         & 00 06 19.52 & $+$20 12 10.5 & 0.026 & 12.6 & 13.9 & S1n  \\
Mrk 590         & 02 14 33.56 & $-$00 46 00.1 & 0.026 & 12.4 & 13.8 & S1.0 \\
Ark 120         & 05 16 11.42 & $-$00 08 59.4 & 0.033 & 12.0 & 13.9 & S1.0 \\
Mrk 79          & 07 42 32.80 & $+$49 48 34.8 & 0.022 & 12.5 & 14.3 & S1.2 \\
PG 0844$+$349   & 08 47 42.40 & $+$34 45 04.0 & 0.064 & 13.4 & 14.5 & S1.0 \\
Mrk 110         & 09 25 12.87 & $+$52 17 10.5 & 0.035 & 13.9 & 16.4 & S1n  \\
NGC 3227        & 10 23 30.58 & $+$19 51 54.2 & 0.004 & 11.3 & 11.8 & S1.5 \\
NGC 4151        & 12 10 32.58 & $+$39 24 20.6 & 0.003 & 10.3 & 11.9 & S1.5 \\
\enddata

\tablecomments{\footnotesize Columns: (1) object name; (2) and (3)
position and (4) redshift from the NASA/IPAC Extragalactic Database
(NED); (5) J magnitude from the Two Micron All Sky Survey
\citep[2MASS;][]{Cutri03}; (6) $V$ magnitude and (7) spectroscopic
classification from the catalogue of \citet{Veron06}, where
S1.0=Seyfert 1, (S1.2, S1.5)=intermediate Seyfert 1, S1n=narrow-line
Seyfert 1, and S3=Seyfert 3 (LINER).}

\end{deluxetable}


\begin{deluxetable}{llrcrrrlrc}
\tabletypesize{\footnotesize} 
\tablecaption{
\label{irtfobslog} 
IRTF Journal of Observations} 
\tablewidth{0pt} 
\tablehead{ 
Object Name & observation & exposure & airmass & \multicolumn{3}{c}{continuum S/N} 
& \multicolumn{3}{c}{telluric standard star} \\
& date & [sec] && J & H & K & name & distance & airmass \\
&&&&&&&& [deg] & \\
(1) & (2) & (3) & (4) & (5) & (6) & (7) & (8) & (9) & (10)
}
\startdata
Mrk 335         & \bf{2006 Jan 10} & 48x120 & 1.422 &  67 & 132 & 269 & HD13869  & 31.8 & 1.033 \\
Mrk 590         & \bf{2006 Jan 09} & 48x120 & 1.128 &  47 &  89 & 160 & HD13936  & 10.2 & 1.333 \\
Ark 120         & \bf{2006 Jan 08} & 56x120 & 1.088 & 226 & 214 & 325 & HD31411  &  7.3 & 1.204 \\
Mrk 79          & \bf{2006 Jan 08} & 56x120 & 1.204 &  68 & 162 & 338 & HD45105  & 12.6 & 1.407 \\
PG 0844$+$349   & \bf{2006 Jan 10} & 48x120 & 1.266 &  40 &  83 & 175 & HD71906  &  4.2 & 1.349 \\
Mrk 110         & \bf{2006 Jan 09} & 56x120 & 1.396 &  31 &  95 & 305 & HD71906  & 17.8 & 1.192 \\
NGC 3227        & \bf{2006 Jan 10} & 16x120 & 1.104 &  57 & 112 & 240 & HD89239  &  7.6 & 1.241 \\
NGC 4151        &     2004 May 23  &  6x120 & 1.130 &  32 &  45 & 121 & HD109615 &  4.9 & 1.100 \\
\enddata

\tablecomments{\footnotesize Columns: (1) object name; (2) observation date
  (bold-face indicates that the corresponding spectrum is shown in
  Fig. \ref{IRTFspectra}); (3) exposure time; (4) average airmass; S/N
  in the continuum over $\sim 100$~\AA~measured at the central
  wavelength of the (5) J, (6) H, and (7) K band; for the telluric
  standard star (8) name, (9) distance from the source, and (10)
  average airmass.}

\end{deluxetable}


\begin{deluxetable}{lllccl}
\tabletypesize{\small} 
\tablecaption{
\label{flwoobslog} 
Tillinghast Journal of Observations} 
\tablewidth{0pt} 
\tablehead{ 
Object Name & IRTF & observation & exposure & airmass & cloud \\ 
& observation & date & & & condition \\
& date && [sec] & & \\
(1) & (2) & (3) & (4) & (5) & (6) 
}
\startdata
Mrk 335         & 2006 Jan 10 & 2005 Dec 29 & 3x300 & 1.09 & photom \\
Mrk 590         & 2006 Jan 09 & 2005 Dec 29 & 2x720 & 1.18 & photom \\
Ark 120         & 2006 Jan 08 & 2005 Dec 29 & 3x480 & 1.18 & photom \\
Mrk 79          & 2006 Jan 08 & 2005 Dec 29 & 3x720 & 1.06 & photom \\
PG 0844$+$349   & 2006 Jan 10 & 2005 Dec 29 & 3x480 & 1.03 & photom \\
Mrk 110         & 2006 Jan 09 & 2006 Jan 05 & 3x540 & 1.07 & clear  \\
NGC 3227        & 2006 Jan 10 & 2006 Jan 05 & 3x330 & 1.02 & clear  \\
NGC 4151        & 2004 May 23 & 2004 May 18 & 2x120 & 1.02 & cirrus \\
\enddata

\tablecomments{\footnotesize Columns: (1) object name; (2) IRTF
observation date (reproduced from Table \ref{irtfobslog}); (3)
Tillinghast UT observation date; (4) exposure time; (5) airmass; and
(6) cloud condition.}

\end{deluxetable}


\begin{deluxetable}{lrclrl}
\tabletypesize{\footnotesize} 
\tablecaption{
\label{ident} 
Observed Near-Infrared Emission Lines} 
\tablewidth{0pt} 
\tablehead{ 
Element & vacuum & transition & term & \multicolumn{2}{c}{\FeII~multiplets} \\ 
& wavelength & type & & wavelength & term \\
& [\AA] &&& [\AA] & \\
(1) & (2) & (3) & (4) & (5) & (6) 
}
\startdata
\HeI         &  7067.2 & E1 & $^3$P$^o$-$^3$S     &&\\
\OII         &  7320.9 & E2 & $^2$D$^o$-$^2$P$^o$ &&\\
             &         &    &                     &  7462.4 & b$^4$D-z$^4$D \\
             &         &    &                     &  7711.7 & b$^4$D-z$^4$D \\
\OI          &  7776.3 & E1 & $^5$S$^o$-$^5$P     &&\\
\FeXI        &  7894.0 & M1 & $^3$P-$^3$P         &&\\
             &         &    &                     &  8157.5 & b$^6$D-v$^4$F \\
             &         &    &                     &  8190.6 & b$^6$D-x$^6$F \\
\enddata

\tablecomments{\footnotesize Columns: (1) ion; (2) vacuum rest
  wavelength; (3) transition type, where E1=allowed transition,
  M1=magnetic dipole forbidden transition, and E2=electric quadrupole
  forbidden transition; (4) transition term; (5) wavelength and (6)
  transition term of neighboring \FeII~multiplets.}

\end{deluxetable}


\begin{deluxetable}{lc|crr|ccc}
\tabletypesize{\small}
\tablecaption{
\label{lines} 
Emission Line Measurements}
\tablewidth{0pt} 
\tablehead{
Object Name & continuum & \multicolumn{3}{c|}{broad component} & \multicolumn{3}{c}{narrow component} \\
& flux & flux & \multicolumn{1}{c}{FWHM} & \multicolumn{1}{c|}{$\Delta$ v} & type & flux & FWHM \\ 
& [erg/s/cm$^2$/\AA] & [erg/s/cm$^2$] & \multicolumn{1}{c}{[km/s]} & 
\multicolumn{1}{c|}{[km/s]} & & [erg/s/cm$^2$] & [km/s] \\
(1) & (2) & (3) & \multicolumn{1}{c}{(4)} & \multicolumn{1}{c|}{(5)} & (6) & (7) & (8)
}
\startdata
\multicolumn{8}{c}{Pa$\alpha$} \\[1mm]
\tableline \\[-2mm]
PDS 456         & 4.73e$-$15 & 4.32e$-$13 & 2022 &         &   &            &     \\
3C 273          & 6.76e$-$15 & 7.66e$-$13 & 2946 & $+$ 328 & e & 3.68e$-$14 & 762 \\
Mrk 876         & 1.03e$-$15 & 1.21e$-$13 & 5505 &         & i & 8.08e$-$15 & 475 \\
HE 1228$+$013   & 9.81e$-$16 & 6.86e$-$14 & 1957 &         & e & 1.38e$-$14 & 560 \\
PG 0844$+$349   & 8.44e$-$16 & 1.12e$-$13 & 2219 & $+$ 402 &   &            &     \\
Mrk 110         & 1.07e$-$15 & 1.83e$-$13 & 1827 &         & e & 1.96e$-$14 & 447 \\
\enddata

\tablecomments{Columns: (1) object name; (2) continuum flux at the
line's vacuum rest wavelength; (3) flux and (4) FWHM of the broad
component (uncorrected for instrumental broadening). Note that the
latter is omitted if the line was measured using the Pa$\alpha$ or
Pa$\beta$ template.; (5) shift of the broad component's center
relative to the expected vacuum rest wavelength. Note that we give
only shifts corresponding to more than two pixels on the detector (for
SpeX spectra $>320$ km s$^{-1}$ and for FAST spectra $>270$ km
s$^{-1}$).; (6) type of subtracted narrow component, where i:
inflected, and e: estimated (see Section \ref{narrow}); (7) flux and
(8) FWHM (uncorrected for instrumental broadening) of the narrow
component.}

\end{deluxetable}


\clearpage
\renewcommand{\topmargin}{0.5cm}

\begin{deluxetable}{lcccccccccccc}
\tabletypesize{\scriptsize}
\rotate
\tablecaption{
\label{iriron}
Iron Emission Lines}
\tablewidth{0pt}
\tablehead{
Object Name & \multicolumn{2}{c}{\FeII~$\lambda 4440-4700$} & \multicolumn{2}{c}{\FeII~$\lambda 8150-8365$} & 
\multicolumn{2}{c}{\FeII~$\lambda 9407$} & \multicolumn{2}{c}{\FeII~$\lambda 9573$} & 
\multicolumn{2}{c}{\FeII~$\lambda 9812$} & \multicolumn{2}{c}{\FeII~$\lambda 9956 + 9998$} \\
& flux & ratio\tablenotemark{\star} & flux & ratio\tablenotemark{\star} & flux & ratio\tablenotemark{\star} & flux & ratio\tablenotemark{\star} & flux & ratio\tablenotemark{\star} & flux & ratio\tablenotemark{\star} \\ 
& [erg/s/cm$^2$] & (10.96\tablenotemark{\star\star}) & [erg/s/cm$^2$] & (1.28) & [erg/s/cm$^2$] & (0.40) & [erg/s/cm$^2$] & (0.29) & [erg/s/cm$^2$] & (0.12) & [erg/s/cm$^2$] & (0.88) \\
(1) & (2) & (3) & (4) & (5) & (6) & (7) & (8) & (9) & (10) & (11) & (12) & (13)
}
\startdata
IRAS 1750$+$508 & 7.48e$-$14 &  7.48 & 5.75e$-$15 &  0.58 &            &      & 1.60e$-$16 & 0.02 &            &      & 3.61e$-$15 & 0.36 \\
PDS 456         & 6.09e$-$13 & 13.36 & 4.94e$-$14 &  1.08 & 1.14e$-$14 & 0.25 & 6.83e$-$15 & 0.15 &            &      & 5.63e$-$14 & 1.23 \\
3C 273          & 7.10e$-$13 & 16.95 & 9.73e$-$14 &  2.32 &            &      &            &      &            &      & 5.38e$-$14 & 1.28 \\
Mrk 876         & 1.37e$-$13 & 21.61 & 1.03e$-$14 &  1.62 &            &      &            &      &            &      & 7.97e$-$15 & 1.26 \\
HE 1228$+$013   & 3.32e$-$13 & 19.53 & 1.96e$-$14 &  1.15 & 6.05e$-$15 & 0.36 &            &      & 3.16e$-$15 & 0.19 & 1.99e$-$14 & 1.17 \\
PG 0844$+$349   & 4.76e$-$13 & 39.34 & 2.31e$-$14 &  1.91 & 6.04e$-$15 & 0.50 &            &      &            &      & 1.45e$-$14 & 1.20 \\
Mrk 110         & 6.88e$-$14 & 22.12 & 2.32e$-$14 &  7.46 &            &      &            &      &            &      & 4.86e$-$15 & 1.56 \\
Mrk 509         & 2.18e$-$13 &  6.23 & 8.27e$-$14 &  2.36 &            &      & 2.54e$-$15 & 0.07 &            &      & 5.12e$-$14 & 1.46 \\
\end{deluxetable}
%
\tablehead{
Object Name & \multicolumn{2}{c}{\FeII~1.0132$+$1.0174 $\mu$m} & \multicolumn{2}{c}{\FeII~1.0174 $\mu$m} & 
\multicolumn{2}{c}{\FeII~1.0491$+$1.0502 $\mu$m} & \multicolumn{2}{c}{\FeII~1.0863 $\mu$m} & 
\multicolumn{2}{c}{\FeII~1.1126 $\mu$m} & FWHM \\
& flux & ratio\tablenotemark{\star} & flux & ratio\tablenotemark{\star} & flux & ratio\tablenotemark{\star} & flux & ratio\tablenotemark{\star} & flux & ratio\tablenotemark{\star} & [km/s] \\ 
& [erg/s/cm$^2$] & (0.30) & [erg/s/cm$^2$] & (0.09) & [erg/s/cm$^2$] & (1.00) & [erg/s/cm$^2$] & (0.56) & [erg/s/cm$^2$] & (0.30) & \\
(1) & (14) & (15) & (16) & (17) & (18) & (19) & (20) & (21) & (22) & (23) & (24)
}
\tablebreak
IRAS 1750$+$508 &            &      &            &      & 1.00e$-$14 & 1.00 &            &      & 2.86e$-$15 & 0.29 & 2335 \\
PDS 456         & 1.15e$-$14 & 0.25 &            &      & 4.56e$-$14 & 1.00 & 2.05e$-$14 & 0.45 & 1.02e$-$14 & 0.22 & 1841 \\
3C 273          & 1.66e$-$14 & 0.40 &            &      & 4.19e$-$14 & 1.00 &            &      & 1.54e$-$14 & 0.37 & 2085 \\
Mrk 876         & 2.31e$-$15 & 0.36 &            &      & 6.34e$-$15 & 1.00 &            &      &            &      & 4813 \\
HE 1228$+$013   & 6.27e$-$15 & 0.37 &            &      & 1.70e$-$14 & 1.00 & 6.84e$-$15 & 0.40 & 1.16e$-$14 & 0.68 & 1273 \\
PG 0844$+$349   & 1.18e$-$15 & 0.10 &            &      & 1.21e$-$14 & 1.00 &            &      & 7.80e$-$15 & 0.64 & 2610 \\
Mrk 110         &            &      & 1.64e$-$15 & 0.53 & 3.11e$-$15 & 1.00 &            &      & 3.92e$-$15 & 1.26 & 1370 \\
Mrk 509         &            &      &            &      & 3.50e$-$14 & 1.00 &            &      & 2.01e$-$14 & 0.57 & 2874 \\
\enddata
\tablenotetext{\star}{observed flux ratio relative to the \FeII~1.0491$+$1.0502 $\mu$m blend (expected theoretical value given below in parenthesis)}
\tablenotetext{\star\star}{excludes \FeII~$\lambda 4668$}
\end{deluxetable}


\clearpage
\pagestyle{empty}

\begin{deluxetable}{lccccccccc}
\tabletypesize{\scriptsize}
\rotate
\tablecaption{
\label{irnarrow}
Near-Infrared Narrow Emission Lines}
\tablewidth{0pt} 
\tablehead{
Object Name & \FeIIf~$\lambda 8617$ & \SIII~$\lambda 9531$ & \CI~$\lambda 9824$ & \CI~$\lambda 9850$ & \SVIII~$\lambda 9911$ & 
\SII~1.0287 $\mu$m & \SII~1.0320 $\mu$m & \SII~1.0336 $\mu$m & \SII~1.0371 $\mu$m \\
& flux & flux & flux & flux & flux & flux & flux & flux & flux \\
& [erg/s/cm$^2$] & [erg/s/cm$^2$] & [erg/s/cm$^2$] & [erg/s/cm$^2$] & [erg/s/cm$^2$] & 
[erg/s/cm$^2$] & [erg/s/cm$^2$] & [erg/s/cm$^2$] & [erg/s/cm$^2$] \\
(1) & (2) & (3) & (4) & (5) & (6) & (7) & (8) & (9) & (10)
}
\startdata
IRAS 1750$+$508 &            & 7.36e$-$15 &            &            &            &            &            &            &            \\
H 1821$+$643    &            & 1.25e$-$14 &            &            &            &            &            &            &            \\
PDS 456         &            & 2.81e$-$15 &            &            & 4.47e$-$16 &            &            & 1.74e$-$15 &            \\
3C 273          &            & 7.26e$-$15 &            &            &            &            &            &            &            \\
Mrk 876         &            & 1.17e$-$14 &            &            &            &            &            &            &            \\
HE 1228$+$013   &            &            &            &            & 9.70e$-$16 &            &            &            &            \\
PG 0844$+$349   &            & 4.57e$-$15 &            & 1.55e$-$15 & 1.47e$-$15 &            & 1.21e$-$15 &            &            \\
Mrk 110         &            & 1.70e$-$14 &            &            & 2.88e$-$15 &            &            &            &            \\
\end{deluxetable}
%
\tablehead{
Object Name & \NI~1.0398 $\mu$m & \NI~1.0407 $\mu$m & \SIX~1.2520 $\mu$m & \FeIIf~1.2567 $\mu$m & \FeIIf~1.2788 $\mu$m & 
\FeIIf~1.3206 $\mu$m & H$_2$~2.1213 $\mu$m & H$_2$~2.2230 $\mu$m & H$_2$~2.2470 $\mu$m \\
& flux & flux & flux & flux & flux & flux & flux & flux & flux \\
& [erg/s/cm$^2$] & [erg/s/cm$^2$] & [erg/s/cm$^2$] & [erg/s/cm$^2$] & [erg/s/cm$^2$] & [erg/s/cm$^2$] & [erg/s/cm$^2$] & [erg/s/cm$^2$] & [erg/s/cm$^2$] \\
(1) & (11) & (12) & (13) & (14) & (15) & (16) & (17) & (18) & (19)
}
\tablebreak
IRAS 1750$+$508 &            &            &            & 7.31e$-$16 &            &            &            &            &            \\
H 1821$+$643    &            &            & 1.86e$-$16 & 1.68e$-$15 &            &            &            &            &            \\
PDS 456         &            & 2.79e$-$16 &            & 1.92e$-$15 &            &            &            &            &            \\
3C 273          &            &            &            &            &            &            &            &            &            \\
Mrk 876         &            &            & 2.09e$-$16 & 6.37e$-$16 & 8.52e$-$16 &            & 2.09e$-$15 &            &            \\
HE 1228$+$013   &            &            &            & 1.12e$-$15 &            &            & 5.46e$-$16 &            &            \\
PG 0844$+$349   &            &            & 6.39e$-$16 &            &            &            & 4.08e$-$16 &            &            \\
Mrk 110         &            &            & 2.95e$-$15 & 4.41e$-$15 &            &            & 1.12e$-$15 & 3.74e$-$16 &            \\
\enddata
\end{deluxetable}

\clearpage
\pagestyle{plaintop}

\begin{deluxetable}{lcccccc}
\tabletypesize{\small}
\tablecaption{
\label{oxygenex} 
Oxygen Emission Line Ratios}
\tablewidth{0pt} 
\tablehead{
Object Name & \OI~$\lambda 8446$ & photon flux & \OI~1.3165 $\mu$m & 
photon flux & \OI~$\lambda 7774$ & photon flux \\
& broad flux & \OI~1.1287/ & broad flux & \OI~1.3165/ & broad flux & 
\OI~$\lambda 7774/$ \\
& [erg/s/cm$^2$] & $\lambda 8446$ & [erg/s/cm$^2$] & $\lambda 8446$ & 
[erg/s/cm$^2$] & $\lambda 8446\tablenotemark{\star}$  \\                  
(1) & (2) & (3) & (4) & (5) & (6) & (7) 
}
\startdata
IRAS 1750$+$508 & 1.47e$-$14 & 0.75 & 4.65e$-$16 & 0.05 & 3.41e$-$15 & 1.07 \\
H 1821$+$643    & 8.00e$-$14 & 0.81 & $-$\tablenotemark{\star\star} & $-$  & 2.25e$-$14 & 1.36 \\
PDS 456         & 1.86e$-$13 & 0.19 & $-$        & $-$  & 3.94e$-$14 & 0.24 \\
3C 273          & 2.43e$-$13 & 0.76 & 1.84e$-$14 & 0.12 & 6.13e$-$14 & 1.93 \\
Mrk 876         & 4.72e$-$14 & 0.48 & $-$        & $-$  & 3.21e$-$14 & 1.21 \\
HE 1228$+$013   & 3.79e$-$14 & 0.84 & 2.15e$-$15 & 0.09 & 1.35e$-$14 & 4.69 \\
PG 0844$+$349   & 3.72e$-$14 & 0.81 &  ?\tablenotemark{\star\star\star} & ?    & 8.84e$-$15 & 1.15 \\
Ark 564         & 4.10e$-$14 & 0.73 &$>$1.83e-16 &$>$0.01 & ?        & ?    \\
H 1934$-$063    & 1.02e$-$13 & 0.70 & 7.78e$-$15 & 0.12 & ?          & ?    \\
\enddata

\tablenotetext{\star}{after subtracting the contribution from fluorescence}
\tablenotetext{\star\star}{$-$: not detected}
\tablenotetext{\star\star\star}{?: not covered by the spectrum}

\end{deluxetable}


\begin{deluxetable}{l|ccc|ccc|ccc|ccc}
\tabletypesize{\scriptsize}
\tablecaption{
\label{Feratios} 
\FeII~Emission Line Ratios}
\tablewidth{0pt} 
\tablehead{
& \multicolumn{3}{c|}{log ($\lambda 4440-4700$/} 
& \multicolumn{3}{c|}{log ($\lambda 4440-4700$/}
& \multicolumn{3}{c|}{log ($\lambda 8150-8365$/} 
& \multicolumn{3}{c}{log ($\lambda 8150-8365$/} \\
& \multicolumn{3}{c|}{1.0491$+$1.0502 $\mu$m)}
& \multicolumn{3}{c|}{1.1126 $\mu$m)} 
& \multicolumn{3}{c|}{1.0491$+$1.0502 $\mu$m)}
& \multicolumn{3}{c}{1.1126 $\mu$m)} \\
& N & mean & $\sigma$\tablenotemark{\star} & N & mean & $\sigma$\tablenotemark{\star} & 
  N & mean & $\sigma$\tablenotemark{\star} & N & mean & $\sigma$\tablenotemark{\star} \\
& (1) & (2) & (3) & (4) & (5) & (6) & (7) & (8) & (9) & (10) & (11) & (12)
}
\startdata
observed &&&&&&&&&&&& \\
all & 
21 & 1.40$\pm$0.06 & 6 & 14 & 1.54$\pm$0.07 & $<$1 & 21 & 0.44$\pm$0.07 & 5 & 14 & 0.61$\pm$0.05 & $<$1 \\
most reliable & 
11 & 1.21$\pm$0.07 & 2 & 11 & 1.49$\pm$0.07 &    1 & 11 & 0.31$\pm$0.09 & 2 & 11 & 0.58$\pm$0.06 & $<$1 \\[0.8mm]
\tableline
predicted && 1.04 &&& 1.56 &&& 0.11 &&& 0.62 & \\
\tableline \\[-2mm]
& \multicolumn{3}{c|}{log ($\lambda 4440-4700$/} 
& \multicolumn{3}{c|}{log ($\lambda 9407$/}
& \multicolumn{3}{c|}{log ($\lambda 9573$/} 
& \multicolumn{3}{c}{log ($\lambda 9956+9998$/} \\
& \multicolumn{3}{c|}{$\lambda 8150-8365$)}
& \multicolumn{3}{c|}{1.0491$+$1.0502 $\mu$m)} 
& \multicolumn{3}{c|}{1.0491$+$1.0502 $\mu$m)}
& \multicolumn{3}{c}{1.0491$+$1.0502 $\mu$m)} \\
& N & mean & $\sigma$\tablenotemark{\star} & N & mean & $\sigma$\tablenotemark{\star} & 
  N & mean & $\sigma$\tablenotemark{\star} & N & mean & $\sigma$\tablenotemark{\star} \\
\enddata

\tablenotetext{\star}{statistical significance of the difference between mean and predicted values}

\end{deluxetable}


\begin{deluxetable}{lcccccc}
\tabletypesize{\small}
\tablecaption{
\label{unidenttab} 
Unidentified Near-Infrared Emission Lines}
\tablewidth{0pt} 
\tablehead{
Object Name & $\lambda 7161$ & $\lambda 7875 + 7896$ & $\lambda 10703 + 10736$ 
& $\lambda 16890$ & $\lambda 16925$ & $\lambda 17004$ \\
& flux & flux & flux & flux & flux & flux \\ 
& [erg/s/cm$^2$] & [erg/s/cm$^2$] & [erg/s/cm$^2$] & [erg/s/cm$^2$] & [erg/s/cm$^2$] & [erg/s/cm$^2$] \\
(1) & (2) & (3) & (4) & (5) & (6) & (7)
}
\startdata
PDS 456         & 1.04e$-$14 & 5.56e$-$14 & 6.52e$-$14 &            & 5.30e$-$14 & 2.42e$-$14 \\
3C 273          &            & 3.57e$-$14 &            &            &            &            \\
HE 1228$+$013   &            & 1.46e$-$14 & 7.50e$-$15 &            &            &            \\
PG 0844$+$349   &            & 2.08e$-$14 & 5.73e$-$15 & 1.01e$-$14 &            &            \\
Mrk 817         &            &            & 6.47e$-$15 & 1.26e$-$14 &            &            \\
Ark 564         &            &            & 1.22e$-$14 & 4.62e$-$15 & 8.42e$-$16 & 2.44e$-$15 \\
Mrk 79          &            &            & 2.90e$-$14 & 4.97e$-$15 & 4.26e$-$15 & 3.54e$-$15 \\
H 1934$-$063    &            &            & 2.13e$-$14 & 5.75e$-$15 & 2.60e$-$15 & 4.85e$-$15 \\
\enddata
\end{deluxetable}

\clearpage

\clearpage

\end{document}